\documentclass[amsmath,amssymb,prl,superscriptaddress,reprint,showpacs,longbibliography]{revtex4-1}
\usepackage{graphicx}
\usepackage{dcolumn}
\usepackage[colorlinks=true,linkcolor=blue,citecolor=blue,urlcolor=blue]{hyperref}
\usepackage[usenames,dvipsnames]{xcolor}
\usepackage{soul}
\usepackage{braket}
\usepackage{bm}
\usepackage{upgreek}
\usepackage{mathtools}
\usepackage{bbm}
\usepackage{multirow}

\renewcommand{\vec}[1]{{\bf #1}}

\begin{document}

\title{Superconductivity from polar fluctuations in multi-orbital systems}


\author{Gaurav Chaudhary}
\email{gc674@cam.ac.uk}
\affiliation{TCM Group, Cavendish Laboratory, University of Cambridge, J. J. Thomson Avenue, Cambridge CB3 0HE, United Kingdom}
\affiliation{Materials Science Division, Argonne National Laboratory, Lemont, IL 60439, USA}

\author{Ivar Martin}
\affiliation{Materials Science Division, Argonne National Laboratory, Lemont, IL 60439, USA}

\date{\today}

\pacs{}
\keywords{}


\begin{abstract} 
Motivated by the superconductivity near paraelectric (PE) to ferroelectric (FE) quantum critical point (QCP) in polar metals, we study polar fluctuation mediated superconductivity in multi-orbital systems. 
The PE to FE QCP is approached  by softening of a transverse optical (TO) phonon that is odd under inversion ($\mathcal{I}$). 
We show that the necessary and sufficient condition for the linear coupling between electron-TO phonon is the presence of multiple orbitals on the Fermi-surface, irrespective of the spin-orbital (SO) coupling, multiple electronic bands, or the vicinity to band crossings.  
We show that the linear coupling to the polar fluctuations (such as TO modes) can generally lead to superconductivity.  
We also show that irrespective of the strong $\vec{k}$ dependence of the effective electron-electron interaction in the BCS channel, quite generally the even-parity channel leads to the highest critical temperature. 
In the presence of additional repulsive electron-electron interactions, an odd-parity spin-triplet channel can become the leading BCS instability.  
Finally, we discuss our results in the context of the superconductivity in $\text{SrTiO}_3$ and $\text{KTaO}_3$ that highlights the importance of the underlying multi-orbital physics if the  superconductivity is mediated by the polar fluctuations.
\end{abstract}

\maketitle

\textit{Introduction}-
Polar metals can be characterized by a symmetry lowering phase transition from a non-polar to polar phase, accompanied by the softening of a polar fluctuation~\cite{Anderson1965,Zhou2020}. 
In the non-polar phase, the fluctuations of the polar order parameter prevent formation of a long range order, instead leading to a  PE behavior with a macroscopic $\mathcal{I}$-symmetry.
As the  QCP is approached,  the  polar fluctuations soften significantly as a precursor of FE order~\cite{FN,Shi2013}.
These polar fluctuations are usually associated with TO phonons. 

Doped $\text{SrTiO}_3$ is a well known example of a polar metal that matches this PE-FE QCP scenario~\cite{Muller1979,Schooley1964}. 
It is also long known to be a superconductor~\cite{Schooley1964}. 
The superconductivity in $\text{SrTiO}_3$ has many puzzling features, including its existence at very low carrier densities and robustness near the PE-FE QCP. 
Recently, the possible role of polar TO phonon modes in stabilizing the superconducting phase has  generated a lot of attention~\cite{Edge2015,Rischau2017,Kanasugi2018,Kedem2018,Kanasugi2019,Tomioka2019,Kozii2019,Marel2019,Enderlein2020,Gastiasoro2022,Kozii2022,Yu2022}. $\text{PbTe}$, $\text{SnTe}$, and $\text{KTaO}_3$ are some other notable examples of polar metals/semimetals that become superconducting at very low densities close to a PE-FE QCP~\cite{Matsushita2006,Hulm1968,Ueno2011, Liu2021}. 
It is thus highly suggestive that the polar fluctuations may play an important role in providing the glue for these superconductors. 

For a single electronic band, the conventional gradient coupling of the electrons to the TO phonon is absent. 
Thus to invoke TO phonon as the mechanism for superconductivity, one requires either assistance from indirect weak effects such as deformation and anisotropy~\cite{Wolfle2018,Ruhman2019,Wolfle2019} or two phonon exchange terms~\cite{Nagi1974,Kiselov2021,Volkov2022}. 
In contrast, it was pointed out that in SO coupled systems, polar fluctuations can directly couple to the electronic system via gradient coupling~\cite{Fu2015,Kozii2015}. 
Its application to the superconductivity in doped $\text{SrTiO}_3$ has been studied~\cite{Kozii2019,Gastiasoro2022,Yu2022,Kozii2022}. 
Recently the direct electron-polar fluctuation coupling was also shown to exist on symmetry grounds in the absence of the SO coupling for multiband systems~\cite{Volkov2020}.

Here, we show that the multi-orbital physics is the necessary and sufficient condition for the gradient coupling of the polar fluctuations to the electronic degrees. 
Such coupling can generically lead to superconductivity even without the SO coupling or multiband effects such as nodal-crossings or multiple Fermi surfaces. 
In the BCS weak coupling paradigm, such electron-TO phonon coupling leads to separate superconducting channels in the same spin and opposite spin sectors.  
In the absence of additional repulsive interactions, the even-parity pairing order in the opposite-spin channel is always the dominant superconducting channel with highest critical temperature ($T_c$). 
We show that the repulsive interactions can lead to higher critical temperature in the odd-parity opposite-spin channel and discuss approximate condition for this to happen.  
Finally, we discuss our main results in context of $\text{SrTiO}_3$ and $\text{KTaO}_3$ superconductors, assuming that the superconductivity near the polar QCP is driven by the gradient coupling to the soft TO mode. 
We argue that the SO coupling only plays a secondary role by modifying the single particle band dispersion. 
The main role is played by the underlying multi-orbital nature of the bands that allow for the gradient coupling to the TO phonons, which is independent of the presence of the SO coupling. 

\begin{figure}[t]
  \includegraphics[width=.45\textwidth]{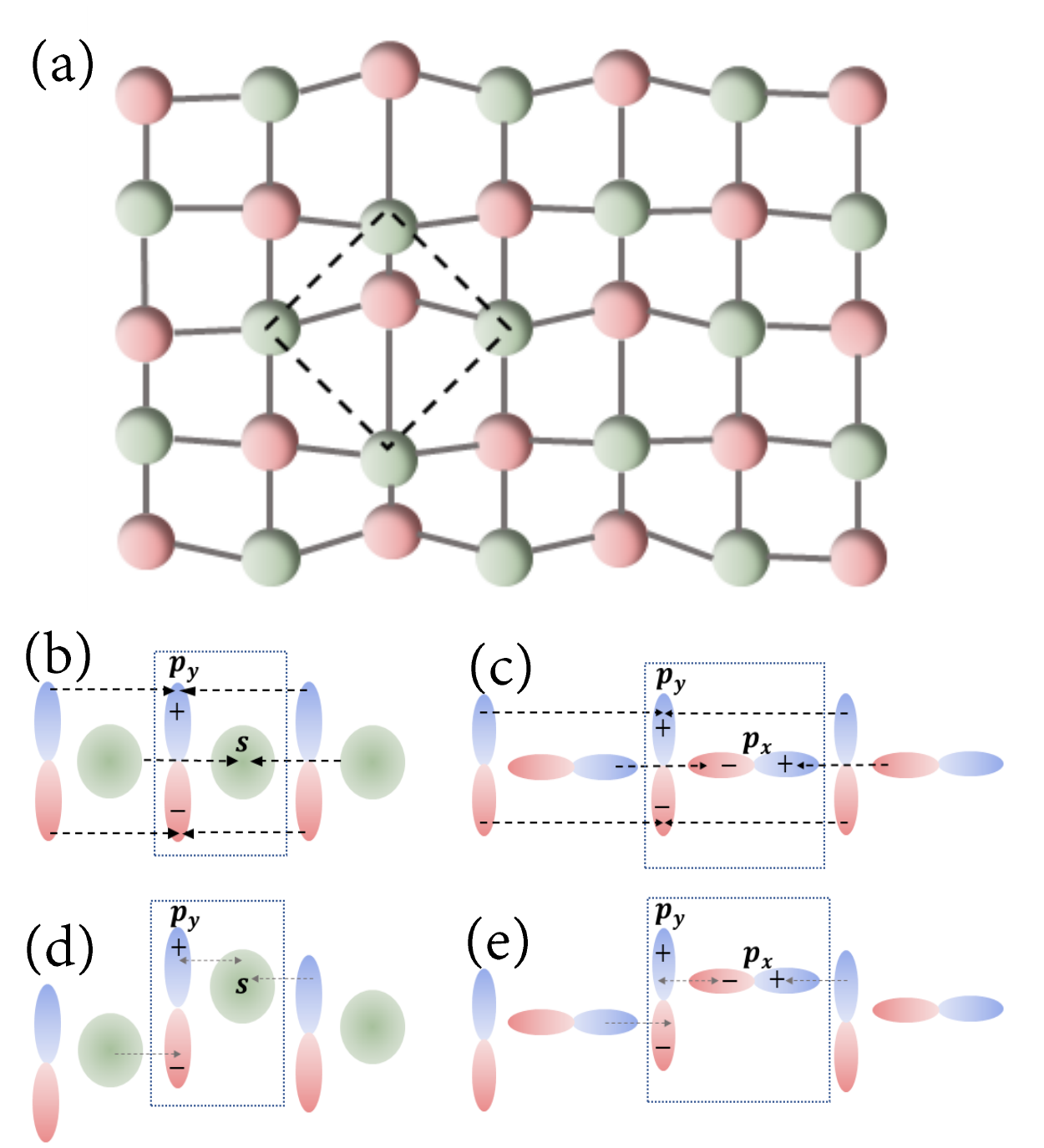}
  \caption{\label{Fig:Orbitals_schematic}
  Polar TO phonon induced inter-orbital hopping channels: (a) shows optical phonon vibration in a two-dimensional bipartite lattice, (b) and (c) respectively show a fluctuationless two orbital system consisting orbitals with opposite transformation under $\mathcal{I}$ ($s$ and $p$ orbital system) and orbitals with same transformation under $\mathcal{I}$ (two different $p$ orbitals). 
  (d) and (e) schematically show their respective TO phonon vibrations. The dynamical breaking of $\mathcal{I}$-symmetry from these vibration induces inter-orbital hopping channels that lead to electron-phonon gradient coupling. 
  The arrow indicate hopping to the $i$-site.
}
\end{figure}

\textit{Model and electron-phonon coupling}-
We consider a general setting of electronic coupling to $\mathcal{I}$-symmetry breaking fluctuations. 
Such setting is studied in previous works in the presence of the SO coupling~\cite{Fu2015}. 
Recently, it was shown that electron coupling to such modes is possible even in the absence of the SO coupling provided that the electronic system is multiband~\cite{Volkov2020}. 
The polar order parameter $\varphi_i$ is odd under inversion, \textit{i.e.} it follows $\mathcal{I}^{-1} \varphi_i \mathcal{I} = - \varphi_i $. 
We consider an electronic system in its PE phase, where the electronic Hamiltonian has both time reversal ($\mathcal{T}$) and the $\mathcal{I}$-symmetry. 
We are interested in the lowest order Yukawa coupling of the type $\int d\vec{k} \varphi_i \hat{O}_i (\vec{k})$ that respects the $\mathcal{I}$-symmetry. 
Since the order parameter is odd under $\mathcal{I}$, we look for the fermionic bilinears $\hat{O}(\vec{k})$ that are also odd under $\mathcal{I}$. 
Ref.~\cite{Volkov2020} considers a spinless two-band system and shows that the symmetry-allowed Yukawa coupling of the polar fluctuation to the electronic Hamiltonian takes a very simple form. 

Spin plays an important role in a superconductor. In the absence of the SO coupling, 
each orbital state is doubly (spin) degenerate, which allows many distinct pairing channels. Thus, we explicitly consider the spinful case given by the electronic Hamiltonian
\begin{align}\label{Eq:H_elec}
    H_e = \sum_{\vec{k}}\sum_s \hat{c}^\dagger_s (\vec{k}) [h_{0,s} (\vec{k}) + \sum^{3}_{i=1} h_{i,s}(\vec{k})\tau_i] \hat{c}_s(\vec{k}),
\end{align}
where $\hat{c}_s(\vec{k}) = (c_{+,s}(\vec{k}),\, c_{-,s}(\vec{k}) )^T$, $s =\uparrow/\downarrow$ is the electron spin with $\hat{z}$ as the spin quantization axis, $\tau_i$ are Pauli matrices and $\pm$ label atomic orbitals. 
The above Hamiltonian is diagonal in spin and seemingly excludes certain SO couplings. 
As we will see, however this is not the case and under appropriate choice of spin quantization axis, the relevant SO couplings are included. 

Under the spinful $\mathcal{T}$ operation, $\mathcal{T}\hat{c}_s(\vec{k})\mathcal{T}^{-1} =  i\sigma_{2,ss'} \hat{c}_{s'}(-\vec{k})$, where $\sigma$ Pauli matrices act in spin space. 
A symmetry under $\mathcal{T}$ along with the hermiticity imposes the constraint $h_{i,s} (\vec{k}) = h_{i,-s}(-\vec{k})$ on the elements of the second-quantized Hamiltonian matrix. 
The inversion acts in the orbital space since the orbitals themselves can change their sign under $\mathcal{I}$. 
Thus we can consider two cases based whether 
 the underlying atomic orbital transform the same of the opposite way under $\mathcal{I}$~\cite{Volkov2020}. 

{\em Opposite orbital parities, Case A}.
When the two orbitals involved have opposite sign under $\mathcal{I}$ [see Fig.~\ref{Fig:Orbitals_schematic} (b) and (d)], \textit{i.e.}, $\mathcal{I} \hat{c}_s(\vec{k}) \mathcal{I}^{-1} = \pm \tau_3 \hat{c}_s(-\vec{k})$, the $\mathcal{I}$ symmetry  enforces that  $h_{0,s} (\vec{k}),\, h_{3,s} (\vec{k}) $ are even functions and $h_{1,s} (\vec{k}),\, h_{2,s} (\vec{k}) $ are odd functions of $\vec{k}$. 
Combining it with the constraints of $\mathcal{T}$, we conclude that a term 
$h_{1,s}$, such that $h_{1,s}(\vec{k}) = -h_{1,s}(-\vec{k})$ and $h_{1,s} = -h_{1,-s}$ is allowed in the fermionic bilinear. 
This term is forbidden for the spinless case~\footnote{The spinless $\mathcal{T}$ is just a complex conjugation, thus $h_1(\vec{k}) = h_1(-\vec{k})$. For case (i), from 
 $\mathcal{I}$, $h_1(\vec{k}) = -h_{1}(-\vec{k})$. Thus $h_1(\vec{k})$ must vanish for spinless case. 
 We can similarly argue that for spinless case (ii), $h_2(\vec{k}) = 0$.
 }
In fact, this term is the general SO coupling that can be introduced in the electronic Hamiltonian that still preserves $\mathcal{I}$ and $\mathcal{T}$ symmetries. 
This can be seen as follows: First, inclusion of $h_{1,s}(\vec{k})$ term neither breaks any degeneracy nor mixes the two spins. 
Its effects are nevertheless quite real, as the spin-rotation symmetry is broken and the two degenerate states at $\vec{k}$ labelled by their spins have different Bloch wavefunctions. 

For comparison, let us consider a general SO term that is off-diagonal in spin. 
The most general SO coupling that still preserves $\mathcal{I}$ and $\mathcal{T}$ is highly constrained to take the form~\cite{SM}
\begin{align}\label{Eq:SO_case1}
    H_{SO} = \sum_{\vec{k}}\sum_{s}\sum_{\tau,\tau'} \lambda(\vec{k})\hat{c}^{\dagger} (\vec{k})\tau_x\sigma_x\hat{c}(\vec{k}),
\end{align}
where $\lambda(\vec{k}) = -\lambda(-\vec{k})$. 
We can make a global rotation of the spin-quantization axis $c_{\rightarrow/\leftarrow} = (c_{\uparrow} \pm c_{\downarrow})/\sqrt{2}$. The electronic Hamiltonian again becomes block-diagonal in spin. Finally, identifying $\lambda(\vec{k}) = h_{1,s}(\vec{k})$, we see that $h_{1,s}(\vec{k})$ is equivalent to inclusion of the SO coupling. 

Having established the general multiorbital electronic Hamiltonian that conserves both $\mathcal{I}$ and $\mathcal{T}$, we can proceed to determine the allowed types of electron phonon coupligns in this case.
The fermionic bilinears that are odd in $\mathcal{I}$ and even in $\mathcal{T}$  can couple to the polar fluctuations. In the small momentum limit, the  electron-phonon Hamiltonian is
\begin{align}\label{Eq:Hq_couple_spin}
    & H^{(1)}_{e-p} = \frac{1}{2}\sum_{\vec{k},\vec{k}'} \sum_{s}  \varphi_i(\vec{k}'-\vec{k}) \hat{c}^{\dagger}_s(\vec{k}')  ([\Gamma^i_1(\vec{k}') + \Gamma^i_1(\vec{k})]\tau_1 \notag\\
    &\hspace{2cm}+s[\Gamma^i_2(\vec{k}') + \Gamma^i_2(\vec{k})]\tau_2  )\hat{c}_s(\vec{k}),
\end{align}
where $\Gamma^i_j (\vec{k}) = \Gamma^i_j(-\vec{k})$ and $s=\pm$ for $\uparrow/\downarrow$. The form of the first term is dictated by the fact that inter-orbital fermion bilinear is odd under $\mathcal{I}$, which forces $\Gamma$ to be even under $\mathcal{I}$.
The extra symmetry-allowed term proportional to $\tau_2$ is only possible in the spinful case. 

{\em Same orbital parities, Case B}.
When both orbitals have the same sign under inversion [see Fig.~\ref{Fig:Orbitals_schematic} (c) and (e)], \textit{i.e.}, $\mathcal{I} \hat{c}_s(\vec{k}) \mathcal{I}^{-1} = \pm \hat{c}_s(-\vec{k})$, the inversion-symmetry enforces that all $h_{i,s} (\vec{k})$ are even functions of $\vec{k}$. Combining it with the constraints of $\mathcal{T}$, we conclude that a term $h_{2,s}$, such that $h_{2,s}(\vec{k}) = h_{2,s}(-\vec{k})$ and $h_{2,s} = -h_{2,-s}$ is allowed in the fermionic bilinear. 
Similar to the previous case, this extra term again encompasses the relevant SO couplings. 
Since the general $\mathcal{I}$ and $\mathcal{T}$ symmetric SO term allowed for this case takes the form~\cite{SM}
\begin{align}\label{Eq:SO_case2}
    H_{SO} = \sum_{\vec{k}}\sum_{s}\sum_{\tau,\tau'} \lambda(\vec{k})\hat{c}^{\dagger} (\vec{k})\tau_x\sigma_y\hat{c}(\vec{k}),
\end{align}
where $\lambda(\vec{k})= \lambda(-\vec{k})$. 
This can be block-diagonalized under global spin rotation: $c_{\rightarrow/\leftarrow} = (c_{\uparrow} \pm ic_{\downarrow})/\sqrt{2}$.
The  Yukawa coupling for this case takes the form 
\begin{align}\label{Eq:Hq_couple}
    & H^{(2)}_{e-p} = \frac{1}{2}
     \sum_{i,\vec{k},\vec{k}'} \varphi_{i}(\vec{k}'-\vec{k}) \hat{c}^{\dagger}(\vec{k}')   ( s[\Gamma^i_1(\vec{k}') + \Gamma^i_1(\vec{k})]\tau_1 \notag\\
     &\hspace{3cm} + [\Gamma^{i}_2(\vec{k}') + \Gamma^{i}_2(\vec{k})] \tau_2 \hat{c}(\vec{k})  ), 
\end{align}
such that $\Gamma^i_{j}(\vec{k}) = -\Gamma^i_j(-\vec{k})$. 
This form is fixed by the fact that inter-orbital fermion bilinears are now even under $\mathcal{I}$, which necessitates $\Gamma$ to be odd function of $\vec k$. 
Similar to the other case, the extra symmetry-allowed term proportional to $\tau_1$ is only possible in the spinfull case.
The sum over $i$ represents a vector-like coupling to the TO phonon. 
From here on we omit the superscript $i$ in $\Gamma(\vec{k})$ and its summation for brevity.  
Above and throughout this work we only consider spin-conserving electron-phonon couplings.

From the form of Eqs.~\ref{Eq:Hq_couple_spin} and~\ref{Eq:Hq_couple}, it is clear that the electron TO phonon coupling is purely inter-orbital irrespective of the presence of the SO coupling. 
Moreover, it is independent of the details (beyond its symmetries) of the single electron Hamiltonian Eq.~\ref{Eq:H_elec}. 
One can tune $h_1$ or $h_2$ such that only one band (per spin) is present at the FS. 
Since because of the finite $h_1$ or $h_2$, the single FS can have significant orbital mixing, there is significant gradient coupling of electron to the TO phonon. 
We note that in Eqs.~\ref{Eq:Hq_couple_spin} and~\ref{Eq:Hq_couple}, the electron-phonon vertex function is comprised of  the sum of the vertex function calculated at momenta before and after the scattering event ($\vec{k}$ and $\vec{k}'$). Strictly, this form applies only if both momenta are small compared to the reciprocal lattice constant.


\textit{Superconductivity}-
In the presence of $\mathcal{I}$ and $\mathcal{T}$ symmetry, a weak coupling BCS superconductor is expected to have condensation of zero momentum Cooper pairs. 
As a result, in the ordered state only the intra-band pairing correlations have a finite order parameter. 
In the single particle band representation the electron-phonon coupling becomes 
\begin{align}\label{Eq:Ham_e_ph_band}
    & H^a_{e-p} = \frac{1}{2}\sum_{\vec{k},\vec{k}'}\sum_s \sum_{j,\eta,\eta'} \varphi(\vec{k}'-\vec{k})  \bar{\Gamma}^{a}_{j,s}(\vec{k},\vec{k}') \tau_{j;\eta,\eta'} \notag\\
    &\hspace{4cm}\times \chi^{\dagger}_{s,\eta}(\vec{k}')\chi_{s,\eta'}(\vec{k}). 
\end{align}
Here $\chi_{s,\eta}(\vec{k})$ is a spin $s$ electron annihilation operator in band $\eta$, $\bar{\Gamma}^{a}$ is the electron-phonon vertex function in band representation(explicit form in the supplementary material [SM]~\cite{SM}), and $\tau_j$ Pauli matrix now acts on the electronic bands. 
Here on we use the label `$a=1,2$' for the two cases of relative orbital symmetries.
Owning to the $\mathcal{T}$, the  matrix elements follow
\begin{align}\label{Eq:elec-pho-vertex-T}
    \bar{\Gamma}^{a}_{j,s}(\vec{k},\vec{k}')\tau_{j;\eta,\eta'} = [\bar{\Gamma}^{a}_{j,-s}(-\vec{k},-\vec{k}')\tau_{j;\eta,\eta'}]^{\ast}.
\end{align}

After integrating out phonons,  we obtain an effective electron-electron interaction Hamiltonian. 
\begin{align}\label{Eq:Ham_ee}
    &H^{a}_{e-e} =  -\sum_{\vec{k},\vec{k'}}\sum_{s,s'} \sum_{i,j,\eta,\eta'} V_{\vec{k}'-\vec{k}} \bar{\Gamma}^{a}_{i,s}(\vec{k},\vec{k}') \bar{\Gamma}^{a}_{j,s'}(-\vec{k},-\vec{k}') \notag\\
    &\hspace{.5cm}\tau_{i;\eta,\eta'}\tau_{j;\eta,\eta'} 
     \chi^{\dagger}_{s,\eta}(\vec{k}')\chi_{s,\eta'}(\vec{k}) \chi^{\dagger}_{s',\eta}(-\vec{k}')\chi_{s',\eta'}(-\vec{k}) 
    .
\end{align}
Here $V_{\vec{q}} = \langle \varphi(\vec{q}) \varphi(-\vec{q})\rangle/4$ is the Bosonic propagator of the polar fluctuations. 
We obtain the BCS Hamiltonian from the interaction Hamiltonian
\begin{align}\label{Eq:H_BCS1}
    & H^{a}_{\text{BCS}} =  - \sum_{\vec{k},\vec{k}'} \sum_{s,s'} \sum_{\eta,\eta'} V^{a}_{\eta,\eta'}(\vec{k},\vec{k'}) \chi^{\dagger}_{s,\eta}(\vec{k}') \chi^{\dagger}_{s',\eta}(-\vec{k}') \notag\\
    &\hspace{4.5cm}\times \chi_{s',\eta'}(-\vec{k}) \chi_{s,\eta'}(\vec{k}),
\end{align}
where 
\begin{align}\label{Eq:V_matrix}
    & V^{a}_{\eta,\eta'}(\vec{k},\vec{k'}) =  \sum_{i,j}V_{\vec{k}-\vec{k}'} \bar{\Gamma}^{a}_{i,s}(\vec{k},\vec{k}') \bar{\Gamma}^{a}_{j,s'}(-\vec{k},-\vec{k}') \notag\\
    &\hspace{4cm}\times\tau_{i;\eta,\eta'}\tau_{j;\eta,\eta'}
\end{align}
are the interaction matrix elements. 

The BCS gap function is a $4\times 4$ matrix in the orbital and the spin space. 
Because the spin is conserved at the electron-phonon vertex, we can separate same spin and opposite spin pairing correlations without any mixing between them. 
Thus reducing the full $4\times 4$ gap function to two separate $2\times 2$ gap function, i.e. a same spin pairing channel and an opposite spin pairing channel. 
The spin state of the same spin pairing channel is triplet. 
In the opposite spin channel both, the spin-singlet and the spin-triplet pairing correlations are present. 
Presence of the SO coupling however, mixes the spin-singlet and spin-triplet pairing correlations in the opposite spin channel, i.e. we cannot write decoupled gap equations for singlet and triplet pairs~\cite{SM}. 

Thus, in the opposite spin channel, we write down a gap equation without singlet/triplet decomposition. In the band basis, after some manipulation, we obtain 
\begin{align}\label{Eq:BCS_gap_band}
    &\Delta^a_{\eta;s,-s}(\vec{k}) = \sum_{\vec{k}',\eta'}V_{\vec{k}'-\vec{k}} |\Upsilon^a_{s;\eta,\eta'}(\vec{k}',\vec{k})|^2 \notag\\
    &\hspace{2cm}\times \frac{\Delta^a_{\eta';,s,-s}(\vec{k}')}{2\xi_{\eta'}(\vec{k'})}\tanh \biggl ( \frac{\xi_{\eta'}(\vec{k}')}{2k_B T}\biggr ),
\end{align}
where
\begin{subequations}\label{Eq:Gap_interaction_matrix}
\begin{align}
    &\Upsilon^a_{s;\eta,\eta} (\vec{k},\vec{k}') =  \bar{\Gamma}^a_{0,s}(\vec{k},\vec{k}') + \eta \bar{\Gamma}^a_{3,s}(\vec{k},\vec{k}') ,  \\
    & \Upsilon^a_{s;\eta,-\eta}(\vec{k},\vec{k}') = (-)^a[\bar{\Gamma}^a_{1,s}(\vec{k},\vec{k}') \mp i \eta \bar{\Gamma}^a_{2,s}(\vec{k},\vec{k}')] . 
\end{align}
\end{subequations}
In obtaining the above form, we have used the $\mathcal{T}$-symmetry constraint of Eq.~\ref{Eq:elec-pho-vertex-T}.
In the limit of soft TO mode, the solutions of the linearized gap equation can be restricted to the F.S. to obtain
\begin{align}\label{Eq:Gap_eigen1}
    & \Delta^a_{\eta;,s,-s}(\vec{k}) \sim \ln\frac{1.14E_D}{k_BT_c} \notag\\
    & \sum'_{\vec{k}'}\sum_{\eta'}    \rho_{\eta'}(0) V_{\vec{k}-\vec{k}'} |\Upsilon^a_{s;\eta,\eta'}(\vec{k},\vec{k'})|^2 \Delta^a_{\eta';,s,-s}(\vec{k}') .
\end{align}
Here $\sum'_{\vec{k}'}$ represents sum over the F.S. momenta. The second line of the gap equation is an eigenvalue equation
\begin{align}\label{Eq:Gap_eigen2}
    \lambda \Delta^a_{\eta,s,-s}(\vec{k}) = \sum'_{\vec{k}'}\sum_{\eta'}    M^a_{\eta,\eta';\vec{k},\vec{k}'} \Delta^a_{\eta';,s,-s}(\vec{k}'), 
\end{align}
where the eigenvalue $\lambda$ is an effective BCS coupling constant and all the entries of the symmetric square matrix $M^a_{\eta,\eta';\vec{k},\vec{k}'} = \rho_{\eta'}(0) V_{\vec{k}-\vec{k}'} |\Upsilon^a_{s;\eta,\eta'}(\vec{k},\vec{k'})|^2$ are real and non-negative. 
It follows from the Perron-Frobenius theorem that a real-positive eigenvalue equal to the spectral radius exists~\cite{Bapat1997}. 
Moreover, the corresponding eigenvector can always be chosen to have non-negative entries. 
These statements respectively prove: (i) a BCS instability with a finite critical temperature $T_c$ exists, and (ii) the BCS gap function corresponding to the maximum $T_c$ has even parity, since it can be chosen to be real and non-negative everywhere.

In the same-spin pairing channel, the intra-band gap function is odd-parity for the Case B and mixed parity for the Case A~\cite{SM}. The BCS gap equations in the same spin channel can be obtained after some manipulations as before 
\begin{align}\label{Eq:BCS_gap_band2}
     &\Delta^a_{\eta;s,s}(\vec{k}) = \sum_{\vec{k}',\eta'}V_{\vec{k}'-\vec{k}} (\Upsilon^a_{s;\eta,\eta'}(\vec{k}',\vec{k}))^2 \notag\\
    &\hspace{2cm}\times \frac{\Delta^a_{\eta';,s,s}(\vec{k}')}{2\xi_{\eta'}(\vec{k'})}\tanh \biggl ( \frac{\xi_{\eta'}(\vec{k}')}{2k_B T}\biggr ). 
\end{align}
Here, we have used the constraint of the $\mathcal{I}$ asymmetry to obtain the above form of the gap equation~\cite{SM}.
Since the entries of the resultant eigenvalue equation are bounded by the corresponding entries of $M^{a}$, we conclude that the spectral radius of this matrix and hence the BCS coupling constant does not exceed the one for Eq.~\ref{Eq:Gap_eigen2}. 
Thus, we have proven that the gap function in the same spin pairing channel does not lead to a $T_c$ larger than that of the even-parity gap function of the opposite spin channel. 

Similar results exclusively in the opposite spin channel were previously proven for the Fr\"olich coupling of a longitudinal phonon to the single electronic band~\cite{Brydon2014}. 
Here, we have generalized these results to vector coupling to the TO phonon mode in both same spin and opposite spin Cooper pair channels. 
Unlike the previous work, here we have also not assumed decoupled gap equations for the spin-singlet and spin-triplet pairing. 
In the presence of the SO coupling, the gap equations for the spin-singlet and spin-triplet (in the opposite spin channel) pairing gap are generally coupled. 
Some previous works have also argued for generally degenerate odd and even-parity gap functions for the polar fluctuation mediated superconductivity~\cite{Kozii2015,Wang2016}. 
We note however that this degeneracy is a consequence of a highly symmetrical electron-phonon vertex function that was considered in those works, which lead to many zero entries of the matrix $M^a$. 
That made the matrix $M^a$ reducible and degenerate for odd- and even-parity solutions. 
More generally, our results show that the $T_c$ of the odd parity solution is bounded by the highest  $T_c$ of the even parity solution. This may change if Coulomb interaction is included.

\textit{Repulsive interactions}- 
We now consider the effect of Coulomb interaction, modelling it as an on-site Hubbard repulsion. 
In the anticipation of zero momentum Cooper pairs, we reduce the Hubbard repulsion to the BCS form, retaining only tthe terms that correspond to  the Cooper pair scattering. 
 For simplicity, we assume the filling such that only one band (per spin) crosses the Fermi energy. We expect that our main results remain valid also  when both bands are present at the Fermi level. 
After projecting on the single particle band basis, the Hubbard Hamiltonian in the BCS channel of the opposite spin pairs takes the form
\begin{align}\label{Eq:Hubbard_band}
    & H^a_{\text{Hub}}  = \sum_{\vec{k},\vec{k}'} \sum_{s,s'}  \bar{U}^a_{s}(\vec{k},\vec{k}')
    \bar{U}^a_{s}(-\vec{k},-\vec{k}') \notag\\
    &\hspace{2cm} \times \chi^{\dagger}_{s}(\vec{k}') \chi^{\dagger}_{-s}(-\vec{k}') \chi_{-s}(-\vec{k}) \chi_{s}(\vec{k}) .
\end{align}
The explicit form of $\bar{U}^a_{s}(\vec{k},\vec{k}')$ is presented in the SM~\cite{SM}. Note that in the equal-spin channel, the local Hubbard repulsion vanishes due to the Pauli exclusion principle.
Further, assuming the $\mathcal{T}$ symmetry,  $\bar{U}^{a}_{s}(\vec{k},\vec{k}') = [\bar{U}^{a}_{-s}(-\vec{k},-\vec{k}')]^{\ast}$. 
Following the same steps as Eq.~\ref{Eq:BCS_gap_band}-\ref{Eq:Gap_eigen2}, we obtain the eigenvalue equation for pairing instability
\begin{align}\label{Eq:Gap_eigen_rep}
    &\lambda \Delta^a_{s,-s}(\vec{k}) = \sum'_{\vec{k}'}    (M^a_{\vec{k},\vec{k}'}- N^a_{\vec{k},\vec{k}'} ) \Delta^a_{s,-s}(\vec{k}'),
\end{align}
where $N^a_{\vec{k},\vec{k}'} = \rho (0) |\bar{U}^a_s(\vec{k},\vec{k'})|^2$. 
We have omitted the band indices $\eta$ in Eqs.~\ref{Eq:Hubbard_band} and~\ref{Eq:Gap_eigen_rep}. 
It is clear that for strong enough repulsive interaction, some of the elements of the  matrix $M^a_{\vec{k},\vec{k}'}- N^a_{\vec{k},\vec{k}'}$ can become negative. 
Hence, the Perron-Frobenius theorem is avoided, raising the possibility of dominant odd-parity channels. 
To explore this possibility, we write $\Delta^a_{s,-s}(\vec{k}) = \text{S}_{\vec{k}} |\Delta^a_{s,-s}(\vec{k})|$, where $\text{S}_{\vec{k}} =\text{Sign}[\Delta^a_{s,-s}(\vec{k})]$, and the gap equation
\begin{align}\label{Eq:Gap_eigen_rep2}
    \lambda |\Delta^a_{s,-s}(\vec{k})| = \sum'_{\vec{k}'} (M^a_{\vec{k},\vec{k}'}- N^a_{\vec{k},\vec{k}'}) \text{S}_{\vec{k}}\text{S}_{\vec{k}'} |\Delta^a_{s,-s}(\vec{k}')|.
\end{align}
We partition all the Fermi momenta $\vec{k}$ into two equally sized mutually exclusive sets $\vec{k}_{+}$ and $\vec{k}_- = -\vec{k}_{+}$. 
For the odd-parity gap the sign of the gap parameter $\Delta^a_{s,-s}(\vec{k})$ is opposite in $\vec{k}_+$ and $\vec{k}_-$. 
We can arrange the diagonalization matrix such that upper half and lower block lie in $\vec{k}_+$ and $\vec{k}_-$ respectively. 
If the repulsive interactions are such that in the off-diagonal blocks (connecting $\vec{k}_+$ to $\vec{k}_-$) their matrix elements predominantly exceed the attractive phonon mediated electronic interactions, while in the diagonal part the electron-phonon interactions exceed the repulsive interactions, the odd-parity gap will be the dominant pairing instability. 
 Roughly the condition for the dominant odd parity BCS instability is 
\begin{align}
    M^a_{\vec{k}_s,\vec{k}'_s} \gtrsim N^a_{\vec{k}_s,\vec{k}'_s}, \quad M^a_{\vec{k}_s,\vec{k}'_{-s}} \lesssim N^a_{\vec{k}_s,\vec{k}'_{-s}}.
\end{align}
The above relations are not strict in the sense that all the matrix elements do not need to follow these relations.


\textit{Discussion}-  
Our main qualitative results can be discussed for the specific cases of $\text{SrTiO}_3$ and $\text{KTaO}_3$ superconductors that lie near the PE-FE QCP and have very different SO coupling strengths.  

Conduction electrons in $\text{SrTiO}_3$ are dominated by the  $t_{2g}$-manifold of the $3d$ orbitals of $\text{Ti}$~\cite{Bistritzer2011,Khalsa2012}. 
To establish a direct relation with our two-orbital model, we consider one pair of orbitals out of the three-fold $t_{2g}$ orbital space at a time. 
Indeed when the cubic symmetry is broken, the $t_{2g}$ manifold is split near the $\Gamma$-point into a degenerate doublet of orbitals and a higher energy orbital.
Consider the doublet formed by the $d_{yz}$ and $d_{zx}$ orbitals. 
Both the $d_{yz}$ and $d_{zx}$ orbital are invariant under $\mathcal{I}$-symmetry. 
Thus it belongs to Case B considered above. 
The atomic SO coupling induced term is still diagonal in the spin when spin quantization axis is chosen along $z$-direction~\cite{Khalsa2012,Gastiasoro2022}, i.e. it is introduced by the terms $h_{2,s} = -h_{2,-s}$. 
All our results were obtained in the presence of this term and apply equally well for the TO phonon coupling in the $d_{yz}$ and $d_{zx}$ orbital doublet.
Even when the career density is so low that only a single band (per spin) is present at the Fermi level, the band can have some orbital mixing even when $\mathcal{I}$ symmetry is preserved, i.e. via symmetry allowed $h_{1,s}(\vec{k})$ in the electronic Hamiltonian. 
Using our main qualitative results, we can conclude that if the repulsive interactions are sufficiently screened in the low density limit, conventional BCS s-wave superconductivity can be mediated by the soft TO phonon as indicated by some experiments~\cite{Thiemann2018, Yoon2021}. 

The case of $\text{KTaO}_3$ follows similar qualitative arguments, except for the conduction electrons are dominated by $t_{2g}$-manifold of $5d$ orbitals of $\text{Ta}$, which have much stronger SO coupling.  
The quantitative differences can nevertheless lead to profound difference in the superconductivity~\cite{Liu2021,Esswein2022,Liu2023a}.
Even though the SO coupling is not critical for the existence of electron-TO phonon coupling, it can still significantly modify the shape of the FS, thus  effecting the Density of states (DOS). 
Furthermore, the strong SO coupling can make $5d$ orbital $t_{2g}$ bands very isotropic near the Brillouin zone center, such that the $3D$ density of state vanishes at small carrier densities (in contrast to small SO case, where individual $t_{2g}$'s remain largely decoupled and with 2D character that ensures finite DOS even at very low densities). 
It is possible that the absence of superconductivity in bulk $\text{KTaO}_3$ is simply due to vanishing density of state of isotropic bulk bands. 
We leave a detailed microscopic analysis that takes into account the actual  values of the SO coupling and electron-phonon couplings of $\text{SrTiO}_3$  and $\text{KTaO}_3$ for future studies. 

In summary, we have shown that necessary and sufficient condition for the gradient coupling of the electrons to the polar-fluctuations (TO phonons) is the presence of multi-orbital physics at the FS. 
The SO coupling that preserves the $\mathcal{I}$ and $\mathcal{T}$ symmetry takes a very restrictive form such that the electronic Hamiltonian can still be block-diagonalized in spin after a simple rotation of the spin quantization axis, even though the SO coupling breaks the spin rotation symmetry. 
Such SO coupling does not break any degeneracy and the single particle bands can be labelled by spin. 
Crucially, the general SO couplings are off-diagonal in the atomic-orbital. 
Thus presence of the SO coupling requires underlying multiple atomic orbitals. 
However, the converse is not true. 
Such electron-TO phonon coupling generically lead to weak coupling BCS instability if the repulsive interactions are sufficiently screened. 
The resultant superconductivity has highest $T_c$ in an even parity opposite spin pairing channel. 
Inclusion of the Coulomb repulsion can favor the odd-parity superconductivity. 
Finally, we have presented approximate conditions on the repulsive interactions for the odd parity pairing to become the dominant BCS instability.



\textit{Acknowledgement}- 
We thank Mike Norman, Anand Bhattacharya, and Jonathan Ruhman for useful discussions.  This work was funded by the Materials Sciences and Engineering Division, Basic Energy Sciences, Office of Science, US DOE.
\bibliography{bibliography}

\begin{thebibliography}{43}%
\makeatletter
\providecommand \@ifxundefined [1]{%
 \@ifx{#1\undefined}
}%
\providecommand \@ifnum [1]{%
 \ifnum #1\expandafter \@firstoftwo
 \else \expandafter \@secondoftwo
 \fi
}%
\providecommand \@ifx [1]{%
 \ifx #1\expandafter \@firstoftwo
 \else \expandafter \@secondoftwo
 \fi
}%
\providecommand \natexlab [1]{#1}%
\providecommand \enquote  [1]{``#1''}%
\providecommand \bibnamefont  [1]{#1}%
\providecommand \bibfnamefont [1]{#1}%
\providecommand \citenamefont [1]{#1}%
\providecommand \href@noop [0]{\@secondoftwo}%
\providecommand \href [0]{\begingroup \@sanitize@url \@href}%
\providecommand \@href[1]{\@@startlink{#1}\@@href}%
\providecommand \@@href[1]{\endgroup#1\@@endlink}%
\providecommand \@sanitize@url [0]{\catcode `\\12\catcode `\$12\catcode
  `\&12\catcode `\#12\catcode `\^12\catcode `\_12\catcode `\%12\relax}%
\providecommand \@@startlink[1]{}%
\providecommand \@@endlink[0]{}%
\providecommand \url  [0]{\begingroup\@sanitize@url \@url }%
\providecommand \@url [1]{\endgroup\@href {#1}{\urlprefix }}%
\providecommand \urlprefix  [0]{URL }%
\providecommand \Eprint [0]{\href }%
\providecommand \doibase [0]{http://dx.doi.org/}%
\providecommand \selectlanguage [0]{\@gobble}%
\providecommand \bibinfo  [0]{\@secondoftwo}%
\providecommand \bibfield  [0]{\@secondoftwo}%
\providecommand \translation [1]{[#1]}%
\providecommand \BibitemOpen [0]{}%
\providecommand \bibitemStop [0]{}%
\providecommand \bibitemNoStop [0]{.\EOS\space}%
\providecommand \EOS [0]{\spacefactor3000\relax}%
\providecommand \BibitemShut  [1]{\csname bibitem#1\endcsname}%
\let\auto@bib@innerbib\@empty
\bibitem [{\citenamefont {Anderson}\ and\ \citenamefont
  {Blount}(1965)}]{Anderson1965}%
  \BibitemOpen
  \bibfield  {author} {\bibinfo {author} {\bibfnamefont {P.~W.}\ \bibnamefont
  {Anderson}}\ and\ \bibinfo {author} {\bibfnamefont {E.~I.}\ \bibnamefont
  {Blount}},\ }\bibfield  {title} {\enquote {\bibinfo {title} {Symmetry
  considerations on martensitic transformations: "ferroelectric" metals?}}\
  }\href {\doibase 10.1103/PhysRevLett.14.217} {\bibfield  {journal} {\bibinfo
  {journal} {Phys. Rev. Lett.}\ }\textbf {\bibinfo {volume} {14}},\ \bibinfo
  {pages} {217} (\bibinfo {year} {1965})}\BibitemShut {NoStop}%
\bibitem [{\citenamefont {Zhou}\ and\ \citenamefont
  {Ariando}(2020)}]{Zhou2020}%
  \BibitemOpen
  \bibfield  {author} {\bibinfo {author} {\bibfnamefont {W.~X.}\ \bibnamefont
  {Zhou}}\ and\ \bibinfo {author} {\bibfnamefont {A.}~\bibnamefont {Ariando}},\
  }\bibfield  {title} {\enquote {\bibinfo {title} {Review on
  ferroelectric/polar metals},}\ }\href {\doibase 10.35848/1347-4065/ab8bbf}
  {\bibfield  {journal} {\bibinfo  {journal} {Jpn. J. Appl. Phys.}\ }\textbf
  {\bibinfo {volume} {59}},\ \bibinfo {pages} {SI0802} (\bibinfo {year}
  {2020})}\BibitemShut {NoStop}%
\bibitem [{FN()}]{FN}%
  \BibitemOpen
  \href@noop {} {}\bibinfo {note} {Strictly metals do not develop
  ferroelectricity due to the screening of the electric field. Polar metals
  thus refer to structural phase that develops a polar axis without a FE charge
  order.}\BibitemShut {Stop}%
\bibitem [{\citenamefont {Shi}\ \emph {et~al.}(2013)\citenamefont {Shi},
  \citenamefont {Guo}, \citenamefont {Wang}, \citenamefont {Princep},
  \citenamefont {Khalyavin}, \citenamefont {Manuel}, \citenamefont {Michiue},
  \citenamefont {Sato}, \citenamefont {Tsuda}, \citenamefont {Yu},
  \citenamefont {Arai}, \citenamefont {Shirako}, \citenamefont {Akaogi},
  \citenamefont {Wang}, \citenamefont {Yamaura},\ and\ \citenamefont
  {Boothroyd}}]{Shi2013}%
  \BibitemOpen
  \bibfield  {author} {\bibinfo {author} {\bibfnamefont {Y.}~\bibnamefont
  {Shi}}, \bibinfo {author} {\bibfnamefont {Y.}~\bibnamefont {Guo}}, \bibinfo
  {author} {\bibfnamefont {X.}~\bibnamefont {Wang}}, \bibinfo {author}
  {\bibfnamefont {A.~J.}\ \bibnamefont {Princep}}, \bibinfo {author}
  {\bibfnamefont {D.}~\bibnamefont {Khalyavin}}, \bibinfo {author}
  {\bibfnamefont {P.}~\bibnamefont {Manuel}}, \bibinfo {author} {\bibfnamefont
  {Y.}~\bibnamefont {Michiue}}, \bibinfo {author} {\bibfnamefont
  {A.}~\bibnamefont {Sato}}, \bibinfo {author} {\bibfnamefont {K.}~\bibnamefont
  {Tsuda}}, \bibinfo {author} {\bibfnamefont {S.}~\bibnamefont {Yu}}, \bibinfo
  {author} {\bibfnamefont {M.}~\bibnamefont {Arai}}, \bibinfo {author}
  {\bibfnamefont {Y.}~\bibnamefont {Shirako}}, \bibinfo {author} {\bibfnamefont
  {M.}~\bibnamefont {Akaogi}}, \bibinfo {author} {\bibfnamefont
  {N.}~\bibnamefont {Wang}}, \bibinfo {author} {\bibfnamefont {K.}~\bibnamefont
  {Yamaura}}, \ and\ \bibinfo {author} {\bibfnamefont {A.~T.}\ \bibnamefont
  {Boothroyd}},\ }\bibfield  {title} {\enquote {\bibinfo {title} {A
  ferroelectric-like structural transition in a metal},}\ }\href {\doibase
  10.1038/nmat3754} {\bibfield  {journal} {\bibinfo  {journal} {Nat. Mater.}\
  }\textbf {\bibinfo {volume} {12}},\ \bibinfo {pages} {1024} (\bibinfo {year}
  {2013})}\BibitemShut {NoStop}%
\bibitem [{\citenamefont {M\"uller}\ and\ \citenamefont
  {Burkard}(1979)}]{Muller1979}%
  \BibitemOpen
  \bibfield  {author} {\bibinfo {author} {\bibfnamefont {K.~A.}\ \bibnamefont
  {M\"uller}}\ and\ \bibinfo {author} {\bibfnamefont {H.}~\bibnamefont
  {Burkard}},\ }\bibfield  {title} {\enquote {\bibinfo {title}
  {$\mathrm{SrTiO}_{3}$: An intrinsic quantum paraelectric below 4 k},}\ }\href
  {\doibase 10.1103/PhysRevB.19.3593} {\bibfield  {journal} {\bibinfo
  {journal} {Phys. Rev. B}\ }\textbf {\bibinfo {volume} {19}},\ \bibinfo
  {pages} {3593} (\bibinfo {year} {1979})}\BibitemShut {NoStop}%
\bibitem [{\citenamefont {Schooley}\ \emph {et~al.}(1964)\citenamefont
  {Schooley}, \citenamefont {Hosler},\ and\ \citenamefont
  {Cohen}}]{Schooley1964}%
  \BibitemOpen
  \bibfield  {author} {\bibinfo {author} {\bibfnamefont {J.~F.}\ \bibnamefont
  {Schooley}}, \bibinfo {author} {\bibfnamefont {W.~R.}\ \bibnamefont
  {Hosler}}, \ and\ \bibinfo {author} {\bibfnamefont {M.~L.}\ \bibnamefont
  {Cohen}},\ }\bibfield  {title} {\enquote {\bibinfo {title} {Superconductivity
  in semiconducting $\mathrm{SrTiO}_{3}$},}\ }\href {\doibase
  10.1103/PhysRevLett.12.474} {\bibfield  {journal} {\bibinfo  {journal} {Phys.
  Rev. Lett.}\ }\textbf {\bibinfo {volume} {12}},\ \bibinfo {pages} {474}
  (\bibinfo {year} {1964})}\BibitemShut {NoStop}%
\bibitem [{\citenamefont {Edge}\ \emph {et~al.}(2015)\citenamefont {Edge},
  \citenamefont {Kedem}, \citenamefont {Aschauer}, \citenamefont {Spaldin},\
  and\ \citenamefont {Balatsky}}]{Edge2015}%
  \BibitemOpen
  \bibfield  {author} {\bibinfo {author} {\bibfnamefont {J.~M.}\ \bibnamefont
  {Edge}}, \bibinfo {author} {\bibfnamefont {Y.}~\bibnamefont {Kedem}},
  \bibinfo {author} {\bibfnamefont {U.}~\bibnamefont {Aschauer}}, \bibinfo
  {author} {\bibfnamefont {N.~A.}\ \bibnamefont {Spaldin}}, \ and\ \bibinfo
  {author} {\bibfnamefont {A.~V.}\ \bibnamefont {Balatsky}},\ }\bibfield
  {title} {\enquote {\bibinfo {title} {Quantum critical origin of the
  superconducting dome in $\mathrm{SrTiO}_{3}$},}\ }\href {\doibase
  10.1103/PhysRevLett.115.247002} {\bibfield  {journal} {\bibinfo  {journal}
  {Phys. Rev. Lett.}\ }\textbf {\bibinfo {volume} {115}},\ \bibinfo {pages}
  {247002} (\bibinfo {year} {2015})}\BibitemShut {NoStop}%
\bibitem [{\citenamefont {Rischau}\ \emph {et~al.}(2017)\citenamefont
  {Rischau}, \citenamefont {Lin}, \citenamefont {Grams}, \citenamefont {Finck},
  \citenamefont {Harms}, \citenamefont {Engelmayer}, \citenamefont {Lorenz},
  \citenamefont {Gallais}, \citenamefont {Fauqu\'e}, \citenamefont
  {Hemberger},\ and\ \citenamefont {Behnia}}]{Rischau2017}%
  \BibitemOpen
  \bibfield  {author} {\bibinfo {author} {\bibfnamefont {C.~W.}\ \bibnamefont
  {Rischau}}, \bibinfo {author} {\bibfnamefont {X.}~\bibnamefont {Lin}},
  \bibinfo {author} {\bibfnamefont {C.~P.}\ \bibnamefont {Grams}}, \bibinfo
  {author} {\bibfnamefont {D.}~\bibnamefont {Finck}}, \bibinfo {author}
  {\bibfnamefont {S.}~\bibnamefont {Harms}}, \bibinfo {author} {\bibfnamefont
  {J.}~\bibnamefont {Engelmayer}}, \bibinfo {author} {\bibfnamefont
  {T.}~\bibnamefont {Lorenz}}, \bibinfo {author} {\bibfnamefont
  {Y.}~\bibnamefont {Gallais}}, \bibinfo {author} {\bibfnamefont
  {B.}~\bibnamefont {Fauqu\'e}}, \bibinfo {author} {\bibfnamefont
  {J.}~\bibnamefont {Hemberger}}, \ and\ \bibinfo {author} {\bibfnamefont
  {K.}~\bibnamefont {Behnia}},\ }\bibfield  {title} {\enquote {\bibinfo {title}
  {A ferroelectric quantum phase transition inside the superconducting dome of
  $\mathrm{Sr}_{1-\mathrm{x}}\mathrm{Ca}_{\mathrm{x}}\mathrm{TiO}_{3-\mathrm{\delta}}$},}\
  }\href {\doibase 10.1038/nphys4085} {\bibfield  {journal} {\bibinfo
  {journal} {Nat. Phys.}\ }\textbf {\bibinfo {volume} {13}},\ \bibinfo {pages}
  {643} (\bibinfo {year} {2017})}\BibitemShut {NoStop}%
\bibitem [{\citenamefont {Kanasugi}\ and\ \citenamefont
  {Yanase}(2018)}]{Kanasugi2018}%
  \BibitemOpen
  \bibfield  {author} {\bibinfo {author} {\bibfnamefont {S.}~\bibnamefont
  {Kanasugi}}\ and\ \bibinfo {author} {\bibfnamefont {Y.}~\bibnamefont
  {Yanase}},\ }\bibfield  {title} {\enquote {\bibinfo {title}
  {Spin-orbit-coupled ferroelectric superconductivity},}\ }\href {\doibase
  10.1103/PhysRevB.98.024521} {\bibfield  {journal} {\bibinfo  {journal} {Phys.
  Rev. B}\ }\textbf {\bibinfo {volume} {98}},\ \bibinfo {pages} {024521}
  (\bibinfo {year} {2018})}\BibitemShut {NoStop}%
\bibitem [{\citenamefont {Kedem}(2018)}]{Kedem2018}%
  \BibitemOpen
  \bibfield  {author} {\bibinfo {author} {\bibfnamefont {Y.}~\bibnamefont
  {Kedem}},\ }\bibfield  {title} {\enquote {\bibinfo {title} {Novel pairing
  mechanism for superconductivity at a vanishing level of doping driven by
  critical ferroelectric modes},}\ }\href {\doibase 10.1103/PhysRevB.98.220505}
  {\bibfield  {journal} {\bibinfo  {journal} {Phys. Rev. B}\ }\textbf {\bibinfo
  {volume} {98}},\ \bibinfo {pages} {220505} (\bibinfo {year}
  {2018})}\BibitemShut {NoStop}%
\bibitem [{\citenamefont {Kanasugi}\ and\ \citenamefont
  {Yanase}(2019)}]{Kanasugi2019}%
  \BibitemOpen
  \bibfield  {author} {\bibinfo {author} {\bibfnamefont {S.}~\bibnamefont
  {Kanasugi}}\ and\ \bibinfo {author} {\bibfnamefont {Y.}~\bibnamefont
  {Yanase}},\ }\bibfield  {title} {\enquote {\bibinfo {title} {Multiorbital
  ferroelectric superconductivity in doped $\mathrm{SrTiO}_{3}$},}\ }\href
  {\doibase 10.1103/PhysRevB.100.094504} {\bibfield  {journal} {\bibinfo
  {journal} {Phys. Rev. B}\ }\textbf {\bibinfo {volume} {100}},\ \bibinfo
  {pages} {094504} (\bibinfo {year} {2019})}\BibitemShut {NoStop}%
\bibitem [{\citenamefont {Tomioka}\ \emph {et~al.}(2019)\citenamefont
  {Tomioka}, \citenamefont {Shirakawa}, \citenamefont {Shibuya},\ and\
  \citenamefont {Inoue}}]{Tomioka2019}%
  \BibitemOpen
  \bibfield  {author} {\bibinfo {author} {\bibfnamefont {Y.}~\bibnamefont
  {Tomioka}}, \bibinfo {author} {\bibfnamefont {N.}~\bibnamefont {Shirakawa}},
  \bibinfo {author} {\bibfnamefont {K.}~\bibnamefont {Shibuya}}, \ and\
  \bibinfo {author} {\bibfnamefont {I.~H.}\ \bibnamefont {Inoue}},\ }\bibfield
  {title} {\enquote {\bibinfo {title} {Enhanced superconductivity close to a
  non-magnetic quantum critical point in electron-doped strontium titanate},}\
  }\href {\doibase 10.1038/s41467-019-08693-1} {\bibfield  {journal} {\bibinfo
  {journal} {Nat. Commun.}\ }\textbf {\bibinfo {volume} {10}},\ \bibinfo
  {pages} {738} (\bibinfo {year} {2019})}\BibitemShut {NoStop}%
\bibitem [{\citenamefont {Kozii}\ \emph {et~al.}(2019)\citenamefont {Kozii},
  \citenamefont {Bi},\ and\ \citenamefont {Ruhman}}]{Kozii2019}%
  \BibitemOpen
  \bibfield  {author} {\bibinfo {author} {\bibfnamefont {V.}~\bibnamefont
  {Kozii}}, \bibinfo {author} {\bibfnamefont {Z.}~\bibnamefont {Bi}}, \ and\
  \bibinfo {author} {\bibfnamefont {J.}~\bibnamefont {Ruhman}},\ }\bibfield
  {title} {\enquote {\bibinfo {title} {Superconductivity near a ferroelectric
  quantum critical point in ultralow-density dirac materials},}\ }\href
  {\doibase 10.1103/PhysRevX.9.031046} {\bibfield  {journal} {\bibinfo
  {journal} {Phys. Rev. X}\ }\textbf {\bibinfo {volume} {9}},\ \bibinfo {pages}
  {031046} (\bibinfo {year} {2019})}\BibitemShut {NoStop}%
\bibitem [{\citenamefont {van~der Marel}\ \emph {et~al.}(2019)\citenamefont
  {van~der Marel}, \citenamefont {Barantani},\ and\ \citenamefont
  {Rischau}}]{Marel2019}%
  \BibitemOpen
  \bibfield  {author} {\bibinfo {author} {\bibfnamefont {D.}~\bibnamefont
  {van~der Marel}}, \bibinfo {author} {\bibfnamefont {F.}~\bibnamefont
  {Barantani}}, \ and\ \bibinfo {author} {\bibfnamefont {C.~W.}\ \bibnamefont
  {Rischau}},\ }\bibfield  {title} {\enquote {\bibinfo {title} {Possible
  mechanism for superconductivity in doped $\mathrm{SrTiO}_{3}$},}\ }\href
  {\doibase 10.1103/PhysRevResearch.1.013003} {\bibfield  {journal} {\bibinfo
  {journal} {Phys. Rev. Res.}\ }\textbf {\bibinfo {volume} {1}},\ \bibinfo
  {pages} {013003} (\bibinfo {year} {2019})}\BibitemShut {NoStop}%
\bibitem [{\citenamefont {Enderlein}\ \emph {et~al.}(2020)\citenamefont
  {Enderlein}, \citenamefont {de~Oliveira}, \citenamefont {Tompsett},
  \citenamefont {Saitovitch}, \citenamefont {Saxena}, \citenamefont
  {Lonzarich},\ and\ \citenamefont {Rowley}}]{Enderlein2020}%
  \BibitemOpen
  \bibfield  {author} {\bibinfo {author} {\bibfnamefont {C.}~\bibnamefont
  {Enderlein}}, \bibinfo {author} {\bibfnamefont {J.~F.}\ \bibnamefont
  {de~Oliveira}}, \bibinfo {author} {\bibfnamefont {D.~A.}\ \bibnamefont
  {Tompsett}}, \bibinfo {author} {\bibfnamefont {E.~B.}\ \bibnamefont
  {Saitovitch}}, \bibinfo {author} {\bibfnamefont {S.~S.}\ \bibnamefont
  {Saxena}}, \bibinfo {author} {\bibfnamefont {G.~G.}\ \bibnamefont
  {Lonzarich}}, \ and\ \bibinfo {author} {\bibfnamefont {S.~E.}\ \bibnamefont
  {Rowley}},\ }\bibfield  {title} {\enquote {\bibinfo {title}
  {Superconductivity mediated by polar modes in ferroelectric metals},}\ }\href
  {\doibase 10.1038/s41467-020-18438-0} {\bibfield  {journal} {\bibinfo
  {journal} {Nat. Commun.}\ }\textbf {\bibinfo {volume} {11}},\ \bibinfo
  {pages} {4852} (\bibinfo {year} {2020})}\BibitemShut {NoStop}%
\bibitem [{\citenamefont {Gastiasoro}\ \emph {et~al.}(2022)\citenamefont
  {Gastiasoro}, \citenamefont {Temperini}, \citenamefont {Barone},\ and\
  \citenamefont {Lorenzana}}]{Gastiasoro2022}%
  \BibitemOpen
  \bibfield  {author} {\bibinfo {author} {\bibfnamefont {M.~N.}\ \bibnamefont
  {Gastiasoro}}, \bibinfo {author} {\bibfnamefont {M.~E.}\ \bibnamefont
  {Temperini}}, \bibinfo {author} {\bibfnamefont {P.}~\bibnamefont {Barone}}, \
  and\ \bibinfo {author} {\bibfnamefont {J.}~\bibnamefont {Lorenzana}},\
  }\bibfield  {title} {\enquote {\bibinfo {title} {Theory of superconductivity
  mediated by rashba coupling in incipient ferroelectrics},}\ }\href {\doibase
  10.1103/PhysRevB.105.224503} {\bibfield  {journal} {\bibinfo  {journal}
  {Phys. Rev. B}\ }\textbf {\bibinfo {volume} {105}},\ \bibinfo {pages}
  {224503} (\bibinfo {year} {2022})}\BibitemShut {NoStop}%
\bibitem [{\citenamefont {Kozii}\ \emph {et~al.}(2022)\citenamefont {Kozii},
  \citenamefont {Klein}, \citenamefont {Fernandes},\ and\ \citenamefont
  {Ruhman}}]{Kozii2022}%
  \BibitemOpen
  \bibfield  {author} {\bibinfo {author} {\bibfnamefont {V.}~\bibnamefont
  {Kozii}}, \bibinfo {author} {\bibfnamefont {A.}~\bibnamefont {Klein}},
  \bibinfo {author} {\bibfnamefont {R.~M.}\ \bibnamefont {Fernandes}}, \ and\
  \bibinfo {author} {\bibfnamefont {J.}~\bibnamefont {Ruhman}},\ }\bibfield
  {title} {\enquote {\bibinfo {title} {Synergetic ferroelectricity and
  superconductivity in zero-density dirac semimetals near quantum
  criticality},}\ }\href {\doibase 10.1103/PhysRevLett.129.237001} {\bibfield
  {journal} {\bibinfo  {journal} {Phys. Rev. Lett.}\ }\textbf {\bibinfo
  {volume} {129}},\ \bibinfo {pages} {237001} (\bibinfo {year}
  {2022})}\BibitemShut {NoStop}%
\bibitem [{\citenamefont {Yu}\ \emph {et~al.}(2022)\citenamefont {Yu},
  \citenamefont {Hwang}, \citenamefont {Raghu},\ and\ \citenamefont
  {Chung}}]{Yu2022}%
  \BibitemOpen
  \bibfield  {author} {\bibinfo {author} {\bibfnamefont {Y.}~\bibnamefont
  {Yu}}, \bibinfo {author} {\bibfnamefont {H.~Y.}\ \bibnamefont {Hwang}},
  \bibinfo {author} {\bibfnamefont {S.}~\bibnamefont {Raghu}}, \ and\ \bibinfo
  {author} {\bibfnamefont {S.~B.}\ \bibnamefont {Chung}},\ }\bibfield  {title}
  {\enquote {\bibinfo {title} {Theory of superconductivity in doped quantum
  paraelectrics},}\ }\href {\doibase 10.1038/s41535-022-00466-2} {\bibfield
  {journal} {\bibinfo  {journal} {npj Quantum Mater.}\ }\textbf {\bibinfo
  {volume} {7}},\ \bibinfo {pages} {63} (\bibinfo {year} {2022})}\BibitemShut
  {NoStop}%
\bibitem [{\citenamefont {Matsushita}\ \emph {et~al.}(2006)\citenamefont
  {Matsushita}, \citenamefont {Wianecki}, \citenamefont {Sommer}, \citenamefont
  {Geballe},\ and\ \citenamefont {Fisher}}]{Matsushita2006}%
  \BibitemOpen
  \bibfield  {author} {\bibinfo {author} {\bibfnamefont {Y.}~\bibnamefont
  {Matsushita}}, \bibinfo {author} {\bibfnamefont {P.~A.}\ \bibnamefont
  {Wianecki}}, \bibinfo {author} {\bibfnamefont {A.~T.}\ \bibnamefont
  {Sommer}}, \bibinfo {author} {\bibfnamefont {T.~H.}\ \bibnamefont {Geballe}},
  \ and\ \bibinfo {author} {\bibfnamefont {I.~R.}\ \bibnamefont {Fisher}},\
  }\bibfield  {title} {\enquote {\bibinfo {title} {Type ii superconducting
  parameters of $\mathrm{Tl}$-doped $\mathrm{PbTe}$ determined from heat
  capacity and electronic transport measurements},}\ }\href {\doibase
  10.1103/PhysRevB.74.134512} {\bibfield  {journal} {\bibinfo  {journal} {Phys.
  Rev. B}\ }\textbf {\bibinfo {volume} {74}},\ \bibinfo {pages} {134512}
  (\bibinfo {year} {2006})}\BibitemShut {NoStop}%
\bibitem [{\citenamefont {Hulm}\ \emph {et~al.}(1968)\citenamefont {Hulm},
  \citenamefont {Jones}, \citenamefont {Deis}, \citenamefont {Fairbank},\ and\
  \citenamefont {Lawless}}]{Hulm1968}%
  \BibitemOpen
  \bibfield  {author} {\bibinfo {author} {\bibfnamefont {J.~K.}\ \bibnamefont
  {Hulm}}, \bibinfo {author} {\bibfnamefont {C.~K.}\ \bibnamefont {Jones}},
  \bibinfo {author} {\bibfnamefont {D.~W.}\ \bibnamefont {Deis}}, \bibinfo
  {author} {\bibfnamefont {H.~A.}\ \bibnamefont {Fairbank}}, \ and\ \bibinfo
  {author} {\bibfnamefont {P.~A.}\ \bibnamefont {Lawless}},\ }\bibfield
  {title} {\enquote {\bibinfo {title} {Superconducting interactions in tin
  telluride},}\ }\href {\doibase 10.1103/PhysRev.169.388} {\bibfield  {journal}
  {\bibinfo  {journal} {Phys. Rev.}\ }\textbf {\bibinfo {volume} {169}},\
  \bibinfo {pages} {388} (\bibinfo {year} {1968})}\BibitemShut {NoStop}%
\bibitem [{\citenamefont {Ueno}\ \emph {et~al.}(2011)\citenamefont {Ueno},
  \citenamefont {Nakamura}, \citenamefont {Shimotani}, \citenamefont {Yuan},
  \citenamefont {Kimura}, \citenamefont {Nojima}, \citenamefont {Aoki},
  \citenamefont {Iwasa},\ and\ \citenamefont {Kawasaki}}]{Ueno2011}%
  \BibitemOpen
  \bibfield  {author} {\bibinfo {author} {\bibfnamefont {K.}~\bibnamefont
  {Ueno}}, \bibinfo {author} {\bibfnamefont {S.}~\bibnamefont {Nakamura}},
  \bibinfo {author} {\bibfnamefont {H.}~\bibnamefont {Shimotani}}, \bibinfo
  {author} {\bibfnamefont {H.~T.}\ \bibnamefont {Yuan}}, \bibinfo {author}
  {\bibfnamefont {N.}~\bibnamefont {Kimura}}, \bibinfo {author} {\bibfnamefont
  {T.}~\bibnamefont {Nojima}}, \bibinfo {author} {\bibfnamefont
  {H.}~\bibnamefont {Aoki}}, \bibinfo {author} {\bibfnamefont {Y.}~\bibnamefont
  {Iwasa}}, \ and\ \bibinfo {author} {\bibfnamefont {M.}~\bibnamefont
  {Kawasaki}},\ }\bibfield  {title} {\enquote {\bibinfo {title} {Discovery of
  superconductivity in $\mathrm{KTaO}_{3}$ by electrostatic carrier doping},}\
  }\href {\doibase 10.1038/nnano.2011.78} {\bibfield  {journal} {\bibinfo
  {journal} {Nat. Nanotechnol.}\ }\textbf {\bibinfo {volume} {6}},\ \bibinfo
  {pages} {408} (\bibinfo {year} {2011})}\BibitemShut {NoStop}%
\bibitem [{\citenamefont {Liu}\ \emph {et~al.}(2021)\citenamefont {Liu},
  \citenamefont {Yan}, \citenamefont {Jin}, \citenamefont {Ma}, \citenamefont
  {Hsiao}, \citenamefont {Lin}, \citenamefont {Bretz-Sullivan}, \citenamefont
  {Zhou}, \citenamefont {Pearson}, \citenamefont {Fisher}, \citenamefont
  {Jiang}, \citenamefont {H.}, \citenamefont {Zuo}, \citenamefont {Wen},
  \citenamefont {Fong}, \citenamefont {Sun}, \citenamefont {Zhou},\ and\
  \citenamefont {Bhattacharya}}]{Liu2021}%
  \BibitemOpen
  \bibfield  {author} {\bibinfo {author} {\bibfnamefont {C.}~\bibnamefont
  {Liu}}, \bibinfo {author} {\bibfnamefont {X.}~\bibnamefont {Yan}}, \bibinfo
  {author} {\bibfnamefont {D.}~\bibnamefont {Jin}}, \bibinfo {author}
  {\bibfnamefont {Y.}~\bibnamefont {Ma}}, \bibinfo {author} {\bibfnamefont
  {H.-W.}\ \bibnamefont {Hsiao}}, \bibinfo {author} {\bibfnamefont
  {Y.}~\bibnamefont {Lin}}, \bibinfo {author} {\bibfnamefont {T.~M.}\
  \bibnamefont {Bretz-Sullivan}}, \bibinfo {author} {\bibfnamefont
  {X.}~\bibnamefont {Zhou}}, \bibinfo {author} {\bibfnamefont {J.}~\bibnamefont
  {Pearson}}, \bibinfo {author} {\bibfnamefont {B.}~\bibnamefont {Fisher}},
  \bibinfo {author} {\bibfnamefont {J.~S.}\ \bibnamefont {Jiang}}, \bibinfo
  {author} {\bibfnamefont {W.}~\bibnamefont {H.}}, \bibinfo {author}
  {\bibfnamefont {J.-M.}\ \bibnamefont {Zuo}}, \bibinfo {author} {\bibfnamefont
  {J.}~\bibnamefont {Wen}}, \bibinfo {author} {\bibfnamefont {D.~D.}\
  \bibnamefont {Fong}}, \bibinfo {author} {\bibfnamefont {J.}~\bibnamefont
  {Sun}}, \bibinfo {author} {\bibfnamefont {H.}~\bibnamefont {Zhou}}, \ and\
  \bibinfo {author} {\bibfnamefont {A.}~\bibnamefont {Bhattacharya}},\
  }\bibfield  {title} {\enquote {\bibinfo {title} {Two-dimensional
  superconductivity and anisotropic transport at $\mathrm{KTaO}_{3}$ (111)
  interfaces},}\ }\href {\doibase 10.1126/science.aba5511} {\bibfield
  {journal} {\bibinfo  {journal} {Science}\ }\textbf {\bibinfo {volume}
  {371}},\ \bibinfo {pages} {716} (\bibinfo {year} {2021})}\BibitemShut
  {NoStop}%
\bibitem [{\citenamefont {W\"olfle}\ and\ \citenamefont
  {Balatsky}(2018)}]{Wolfle2018}%
  \BibitemOpen
  \bibfield  {author} {\bibinfo {author} {\bibfnamefont {P.}~\bibnamefont
  {W\"olfle}}\ and\ \bibinfo {author} {\bibfnamefont {A.~V.}\ \bibnamefont
  {Balatsky}},\ }\bibfield  {title} {\enquote {\bibinfo {title}
  {Superconductivity at low density near a ferroelectric quantum critical
  point: Doped $\mathrm{SrTiO}_{3}$},}\ }\href {\doibase
  10.1103/PhysRevB.98.104505} {\bibfield  {journal} {\bibinfo  {journal} {Phys.
  Rev. B}\ }\textbf {\bibinfo {volume} {98}},\ \bibinfo {pages} {104505}
  (\bibinfo {year} {2018})}\BibitemShut {NoStop}%
\bibitem [{\citenamefont {Ruhman}\ and\ \citenamefont
  {Lee}(2019)}]{Ruhman2019}%
  \BibitemOpen
  \bibfield  {author} {\bibinfo {author} {\bibfnamefont {J.}~\bibnamefont
  {Ruhman}}\ and\ \bibinfo {author} {\bibfnamefont {P.~A.}\ \bibnamefont
  {Lee}},\ }\bibfield  {title} {\enquote {\bibinfo {title} {Comment on
  ``superconductivity at low density near a ferroelectric quantum critical
  point: Doped $\mathrm{SrTiO}_{3}$''},}\ }\href {\doibase
  10.1103/PhysRevB.100.226501} {\bibfield  {journal} {\bibinfo  {journal}
  {Phys. Rev. B}\ }\textbf {\bibinfo {volume} {100}},\ \bibinfo {pages}
  {226501} (\bibinfo {year} {2019})}\BibitemShut {NoStop}%
\bibitem [{\citenamefont {W\"olfle}\ and\ \citenamefont
  {Balatsky}(2019)}]{Wolfle2019}%
  \BibitemOpen
  \bibfield  {author} {\bibinfo {author} {\bibfnamefont {P.}~\bibnamefont
  {W\"olfle}}\ and\ \bibinfo {author} {\bibfnamefont {A.}~\bibnamefont
  {Balatsky}},\ }\bibfield  {title} {\enquote {\bibinfo {title} {Reply to
  ``comment on `superconductivity at low density near a ferroelectric quantum
  critical point: Doped $\mathrm{SrTiO}_{3}$'''},}\ }\href {\doibase
  10.1103/PhysRevB.100.226502} {\bibfield  {journal} {\bibinfo  {journal}
  {Phys. Rev. B}\ }\textbf {\bibinfo {volume} {100}},\ \bibinfo {pages}
  {226502} (\bibinfo {year} {2019})}\BibitemShut {NoStop}%
\bibitem [{\citenamefont {Ngai}(1974)}]{Nagi1974}%
  \BibitemOpen
  \bibfield  {author} {\bibinfo {author} {\bibfnamefont {K.~L.}\ \bibnamefont
  {Ngai}},\ }\bibfield  {title} {\enquote {\bibinfo {title} {Two-phonon
  deformation potential and superconductivity in degenerate semiconductors},}\
  }\href {\doibase 10.1103/PhysRevLett.32.215} {\bibfield  {journal} {\bibinfo
  {journal} {Phys. Rev. Lett.}\ }\textbf {\bibinfo {volume} {32}},\ \bibinfo
  {pages} {215} (\bibinfo {year} {1974})}\BibitemShut {NoStop}%
\bibitem [{\citenamefont {Kiselov}\ and\ \citenamefont
  {Feigel'man}(2021)}]{Kiselov2021}%
  \BibitemOpen
  \bibfield  {author} {\bibinfo {author} {\bibfnamefont {D.~E.}\ \bibnamefont
  {Kiselov}}\ and\ \bibinfo {author} {\bibfnamefont {M.~V.}\ \bibnamefont
  {Feigel'man}},\ }\bibfield  {title} {\enquote {\bibinfo {title} {Theory of
  superconductivity due to ngai's mechanism in lightly doped
  $\mathrm{SrTiO}_{3}$},}\ }\href {\doibase 10.1103/PhysRevB.104.L220506}
  {\bibfield  {journal} {\bibinfo  {journal} {Phys. Rev. B}\ }\textbf {\bibinfo
  {volume} {104}},\ \bibinfo {pages} {L220506} (\bibinfo {year}
  {2021})}\BibitemShut {NoStop}%
\bibitem [{\citenamefont {Volkov}\ \emph {et~al.}(2022)\citenamefont {Volkov},
  \citenamefont {Chandra},\ and\ \citenamefont {Coleman}}]{Volkov2022}%
  \BibitemOpen
  \bibfield  {author} {\bibinfo {author} {\bibfnamefont {P.~A.}\ \bibnamefont
  {Volkov}}, \bibinfo {author} {\bibfnamefont {P.}~\bibnamefont {Chandra}}, \
  and\ \bibinfo {author} {\bibfnamefont {P.}~\bibnamefont {Coleman}},\
  }\bibfield  {title} {\enquote {\bibinfo {title} {Superconductivity from
  energy fluctuations in dilute quantum critical polar metals},}\ }\href
  {\doibase 10.1038/s41467-022-32303-2} {\bibfield  {journal} {\bibinfo
  {journal} {Nat. Commun.}\ }\textbf {\bibinfo {volume} {13}},\ \bibinfo
  {pages} {4599} (\bibinfo {year} {2022})}\BibitemShut {NoStop}%
\bibitem [{\citenamefont {Fu}(2015)}]{Fu2015}%
  \BibitemOpen
  \bibfield  {author} {\bibinfo {author} {\bibfnamefont {L.}~\bibnamefont
  {Fu}},\ }\bibfield  {title} {\enquote {\bibinfo {title} {Parity-breaking
  phases of spin-orbit-coupled metals with gyrotropic, ferroelectric, and
  multipolar orders},}\ }\href {\doibase 10.1103/PhysRevLett.115.026401}
  {\bibfield  {journal} {\bibinfo  {journal} {Phys. Rev. Lett.}\ }\textbf
  {\bibinfo {volume} {115}},\ \bibinfo {pages} {026401} (\bibinfo {year}
  {2015})}\BibitemShut {NoStop}%
\bibitem [{\citenamefont {Kozii}\ and\ \citenamefont {Fu}(2015)}]{Kozii2015}%
  \BibitemOpen
  \bibfield  {author} {\bibinfo {author} {\bibfnamefont {V.}~\bibnamefont
  {Kozii}}\ and\ \bibinfo {author} {\bibfnamefont {L.}~\bibnamefont {Fu}},\
  }\bibfield  {title} {\enquote {\bibinfo {title} {Odd-parity superconductivity
  in the vicinity of inversion symmetry breaking in spin-orbit-coupled
  systems},}\ }\href {\doibase 10.1103/PhysRevLett.115.207002} {\bibfield
  {journal} {\bibinfo  {journal} {Phys. Rev. Lett.}\ }\textbf {\bibinfo
  {volume} {115}},\ \bibinfo {pages} {207002} (\bibinfo {year}
  {2015})}\BibitemShut {NoStop}%
\bibitem [{\citenamefont {Volkov}\ and\ \citenamefont
  {Chandra}(2020)}]{Volkov2020}%
  \BibitemOpen
  \bibfield  {author} {\bibinfo {author} {\bibfnamefont {P.~A.}\ \bibnamefont
  {Volkov}}\ and\ \bibinfo {author} {\bibfnamefont {P.}~\bibnamefont
  {Chandra}},\ }\bibfield  {title} {\enquote {\bibinfo {title} {Multiband
  quantum criticality of polar metals},}\ }\href {\doibase
  10.1103/PhysRevLett.124.237601} {\bibfield  {journal} {\bibinfo  {journal}
  {Phys. Rev. Lett.}\ }\textbf {\bibinfo {volume} {124}},\ \bibinfo {pages}
  {237601} (\bibinfo {year} {2020})}\BibitemShut {NoStop}%
\bibitem [{Note1()}]{Note1}%
  \BibitemOpen
  \bibinfo {note} {The spinless $\protect \mathcal {T}$ is just a complex
  conjugation, thus $h_1({\protect \bf k}) = h_1(-{\protect \bf k})$. For case
  (i), from $\protect \mathcal {I}$, $h_1({\protect \bf k}) = -h_{1}(-{\protect
  \bf k})$. Thus $h_1({\protect \bf k})$ must vanish for spinless case. We can
  similarly argue that for spinless case (ii), $h_2({\protect \bf k}) =
  0$.}\BibitemShut {Stop}%
\bibitem [{SM()}]{SM}%
  \BibitemOpen
  \href@noop {} {}\bibinfo {note} {Supplementary material}\BibitemShut
  {NoStop}%
\bibitem [{\citenamefont {Bapat}\ and\ \citenamefont
  {Raghavan}(1997)}]{Bapat1997}%
  \BibitemOpen
  \bibfield  {author} {\bibinfo {author} {\bibfnamefont {R.~B.}\ \bibnamefont
  {Bapat}}\ and\ \bibinfo {author} {\bibfnamefont {T.~E.~S.}\ \bibnamefont
  {Raghavan}},\ }\href {\doibase 10.1017/CBO9780511529979} {\emph {\bibinfo
  {title} {Nonnegative Matrices and Applications}}},\ Encyclopedia of
  Mathematics and its Applications\ (\bibinfo  {publisher} {Cambridge
  University Press},\ \bibinfo {year} {1997})\BibitemShut {NoStop}%
\bibitem [{\citenamefont {Brydon}\ \emph {et~al.}(2014)\citenamefont {Brydon},
  \citenamefont {Sarma}, \citenamefont {Hui},\ and\ \citenamefont
  {Sau}}]{Brydon2014}%
  \BibitemOpen
  \bibfield  {author} {\bibinfo {author} {\bibfnamefont {P.~M.~R.}\
  \bibnamefont {Brydon}}, \bibinfo {author} {\bibfnamefont {S.~Das}\
  \bibnamefont {Sarma}}, \bibinfo {author} {\bibfnamefont {H.-Y.}\ \bibnamefont
  {Hui}}, \ and\ \bibinfo {author} {\bibfnamefont {J.~D.}\ \bibnamefont
  {Sau}},\ }\bibfield  {title} {\enquote {\bibinfo {title} {Odd-parity
  superconductivity from phonon-mediated pairing: Application to
  $\mathrm{Cu}_{x}\mathrm{Bi}_{2}\mathrm{Se}_{3}$},}\ }\href {\doibase
  10.1103/PhysRevB.90.184512} {\bibfield  {journal} {\bibinfo  {journal} {Phys.
  Rev. B}\ }\textbf {\bibinfo {volume} {90}},\ \bibinfo {pages} {184512}
  (\bibinfo {year} {2014})}\BibitemShut {NoStop}%
\bibitem [{\citenamefont {Wang}\ \emph {et~al.}(2016)\citenamefont {Wang},
  \citenamefont {Cho}, \citenamefont {Hughes},\ and\ \citenamefont
  {Fradkin}}]{Wang2016}%
  \BibitemOpen
  \bibfield  {author} {\bibinfo {author} {\bibfnamefont {Y.}~\bibnamefont
  {Wang}}, \bibinfo {author} {\bibfnamefont {G.~Y.}\ \bibnamefont {Cho}},
  \bibinfo {author} {\bibfnamefont {T.~L.}\ \bibnamefont {Hughes}}, \ and\
  \bibinfo {author} {\bibfnamefont {E.}~\bibnamefont {Fradkin}},\ }\bibfield
  {title} {\enquote {\bibinfo {title} {Topological superconducting phases from
  inversion symmetry breaking order in spin-orbit-coupled systems},}\ }\href
  {\doibase 10.1103/PhysRevB.93.134512} {\bibfield  {journal} {\bibinfo
  {journal} {Phys. Rev. B}\ }\textbf {\bibinfo {volume} {93}},\ \bibinfo
  {pages} {134512} (\bibinfo {year} {2016})}\BibitemShut {NoStop}%
\bibitem [{\citenamefont {Bistritzer}\ \emph {et~al.}(2011)\citenamefont
  {Bistritzer}, \citenamefont {Khalsa},\ and\ \citenamefont
  {MacDonald}}]{Bistritzer2011}%
  \BibitemOpen
  \bibfield  {author} {\bibinfo {author} {\bibfnamefont {R.}~\bibnamefont
  {Bistritzer}}, \bibinfo {author} {\bibfnamefont {G.}~\bibnamefont {Khalsa}},
  \ and\ \bibinfo {author} {\bibfnamefont {A.~H.}\ \bibnamefont {MacDonald}},\
  }\bibfield  {title} {\enquote {\bibinfo {title} {Electronic structure of
  doped $\mathrm{d}^{0}$ perovskite semiconductors},}\ }\href {\doibase
  10.1103/PhysRevB.83.115114} {\bibfield  {journal} {\bibinfo  {journal} {Phys.
  Rev. B}\ }\textbf {\bibinfo {volume} {83}},\ \bibinfo {pages} {115114}
  (\bibinfo {year} {2011})}\BibitemShut {NoStop}%
\bibitem [{\citenamefont {Khalsa}\ and\ \citenamefont
  {MacDonald}(2012)}]{Khalsa2012}%
  \BibitemOpen
  \bibfield  {author} {\bibinfo {author} {\bibfnamefont {G.}~\bibnamefont
  {Khalsa}}\ and\ \bibinfo {author} {\bibfnamefont {A.~H.}\ \bibnamefont
  {MacDonald}},\ }\bibfield  {title} {\enquote {\bibinfo {title} {Theory of the
  $\mathrm{SrTiO}_{3}$ surface state two-dimensional electron gas},}\ }\href
  {\doibase 10.1103/PhysRevB.86.125121} {\bibfield  {journal} {\bibinfo
  {journal} {Phys. Rev. B}\ }\textbf {\bibinfo {volume} {86}},\ \bibinfo
  {pages} {125121} (\bibinfo {year} {2012})}\BibitemShut {NoStop}%
\bibitem [{\citenamefont {Thiemann}\ \emph {et~al.}(2018)\citenamefont
  {Thiemann}, \citenamefont {Beutel}, \citenamefont {Dressel}, \citenamefont
  {Lee-Hone}, \citenamefont {Broun}, \citenamefont {Fillis-Tsirakis},
  \citenamefont {Boschker}, \citenamefont {Mannhart},\ and\ \citenamefont
  {Scheffler}}]{Thiemann2018}%
  \BibitemOpen
  \bibfield  {author} {\bibinfo {author} {\bibfnamefont {M.}~\bibnamefont
  {Thiemann}}, \bibinfo {author} {\bibfnamefont {M.~H.}\ \bibnamefont
  {Beutel}}, \bibinfo {author} {\bibfnamefont {M.}~\bibnamefont {Dressel}},
  \bibinfo {author} {\bibfnamefont {N.~R.}\ \bibnamefont {Lee-Hone}}, \bibinfo
  {author} {\bibfnamefont {D.~M.}\ \bibnamefont {Broun}}, \bibinfo {author}
  {\bibfnamefont {E.}~\bibnamefont {Fillis-Tsirakis}}, \bibinfo {author}
  {\bibfnamefont {H.}~\bibnamefont {Boschker}}, \bibinfo {author}
  {\bibfnamefont {J.}~\bibnamefont {Mannhart}}, \ and\ \bibinfo {author}
  {\bibfnamefont {M.}~\bibnamefont {Scheffler}},\ }\bibfield  {title} {\enquote
  {\bibinfo {title} {Single-gap superconductivity and dome of superfluid
  density in nb-doped $\mathrm{SrTiO}_{3}$},}\ }\href {\doibase
  10.1103/PhysRevLett.120.237002} {\bibfield  {journal} {\bibinfo  {journal}
  {Phys. Rev. Lett.}\ }\textbf {\bibinfo {volume} {120}},\ \bibinfo {pages}
  {237002} (\bibinfo {year} {2018})}\BibitemShut {NoStop}%
\bibitem [{\citenamefont {Yoon}\ \emph {et~al.}()\citenamefont {Yoon},
  \citenamefont {Swartz}, \citenamefont {Harvey}, \citenamefont {Inoue},
  \citenamefont {Hikita}, \citenamefont {Yu}, \citenamefont {Chung},
  \citenamefont {Raghu},\ and\ \citenamefont {Hwang}}]{Yoon2021}%
  \BibitemOpen
  \bibfield  {author} {\bibinfo {author} {\bibfnamefont {H.}~\bibnamefont
  {Yoon}}, \bibinfo {author} {\bibfnamefont {A.~G.}\ \bibnamefont {Swartz}},
  \bibinfo {author} {\bibfnamefont {S.~P.}\ \bibnamefont {Harvey}}, \bibinfo
  {author} {\bibfnamefont {H.}~\bibnamefont {Inoue}}, \bibinfo {author}
  {\bibfnamefont {Y.}~\bibnamefont {Hikita}}, \bibinfo {author} {\bibfnamefont
  {Y.}~\bibnamefont {Yu}}, \bibinfo {author} {\bibfnamefont {S.~B.}\
  \bibnamefont {Chung}}, \bibinfo {author} {\bibfnamefont {S.}~\bibnamefont
  {Raghu}}, \ and\ \bibinfo {author} {\bibfnamefont {H.~Y.}\ \bibnamefont
  {Hwang}},\ }\bibfield  {title} {\enquote {\bibinfo {title} {Low-density
  superconductivity in $\mathrm{SrTiO}_{3}$ bounded by the adiabatic
  criterion},}\ }\href@noop {} {\ }\Eprint
  {http://arxiv.org/abs/arXiv:2106.10802} {arXiv:2106.10802} \BibitemShut
  {NoStop}%
\bibitem [{\citenamefont {Esswein}\ and\ \citenamefont
  {Spaldin}()}]{Esswein2022}%
  \BibitemOpen
  \bibfield  {author} {\bibinfo {author} {\bibfnamefont {T.}~\bibnamefont
  {Esswein}}\ and\ \bibinfo {author} {\bibfnamefont {N.~A.}\ \bibnamefont
  {Spaldin}},\ }\bibfield  {title} {\enquote {\bibinfo {title}
  {First-principles calculation of electron-phonon coupling in doped ktao3},}\
  }\href@noop {} {\ }\Eprint {http://arxiv.org/abs/arXiv:2210.14113}
  {arXiv:2210.14113} \BibitemShut {NoStop}%
\bibitem [{\citenamefont {Liu}\ \emph {et~al.}(2023)\citenamefont {Liu},
  \citenamefont {Zhou}, \citenamefont {Hong}, \citenamefont {Fisher},
  \citenamefont {Zheng}, \citenamefont {Pearson}, \citenamefont {Jiang},
  \citenamefont {Jin}, \citenamefont {Norman},\ and\ \citenamefont
  {Bhattacharya}}]{Liu2023a}%
  \BibitemOpen
  \bibfield  {author} {\bibinfo {author} {\bibfnamefont {C.}~\bibnamefont
  {Liu}}, \bibinfo {author} {\bibfnamefont {X.}~\bibnamefont {Zhou}}, \bibinfo
  {author} {\bibfnamefont {D.}~\bibnamefont {Hong}}, \bibinfo {author}
  {\bibfnamefont {B.}~\bibnamefont {Fisher}}, \bibinfo {author} {\bibfnamefont
  {H.}~\bibnamefont {Zheng}}, \bibinfo {author} {\bibfnamefont
  {J.}~\bibnamefont {Pearson}}, \bibinfo {author} {\bibfnamefont {J.~S.}\
  \bibnamefont {Jiang}}, \bibinfo {author} {\bibfnamefont {D.}~\bibnamefont
  {Jin}}, \bibinfo {author} {\bibfnamefont {M.~R.}\ \bibnamefont {Norman}}, \
  and\ \bibinfo {author} {\bibfnamefont {A.}~\bibnamefont {Bhattacharya}},\
  }\bibfield  {title} {\enquote {\bibinfo {title} {Tunable superconductivity
  and its origin at ktao3 interfaces},}\ }\href {\doibase
  10.1038/s41467-023-36309-2} {\bibfield  {journal} {\bibinfo  {journal} {Nat.
  Commun.}\ }\textbf {\bibinfo {volume} {14}},\ \bibinfo {pages} {951}
  (\bibinfo {year} {2023})}\BibitemShut {NoStop}%
\bibitem [{\citenamefont {Sigrist}\ and\ \citenamefont
  {Ueda}(1991)}]{Sigrist1991}%
  \BibitemOpen
  \bibfield  {author} {\bibinfo {author} {\bibfnamefont {M.}~\bibnamefont
  {Sigrist}}\ and\ \bibinfo {author} {\bibfnamefont {K.}~\bibnamefont {Ueda}},\
  }\bibfield  {title} {\enquote {\bibinfo {title} {Phenomenological theory of
  unconventional superconductivity},}\ }\href {\doibase
  10.1103/RevModPhys.63.239} {\bibfield  {journal} {\bibinfo  {journal} {Rev.
  Mod. Phys.}\ }\textbf {\bibinfo {volume} {63}},\ \bibinfo {pages} {239}
  (\bibinfo {year} {1991})}\BibitemShut {NoStop}%
\end{thebibliography}%

\newpage

\begin{widetext}

\subsection{Electron coupling to the transverse optical phonon}
In the main text, we wrote down general symmetry based Yukawa couplings to the TO phonon in a multi-orbital systems. 
Here, we microscopically derive the electron-polar TO phonon coupling for a minimal model and show that it follows the symmetry based results of the main text. 
Since we are interested in a polar mode, at minimal we need to consider a diatomic unit cell that can facilitate a local polarization. 
We associate different orbitals with the different atomic sites in the unit cell. 
For definiteness, we restrict to two dimensional bipartite square lattice, where one sub-lattice hosts a $p_y$ orbital and the other sublattice hosts either an $s$ or a $p_x$ orbital. 
A schematic of the 2D lattice under consideration is presented in Fig.~\ref{Fig:Orbitals_schematic} (a) of the main text. 
The generalization to three dimensions or different orbitals and unit cells is straightforward. 
Let's assume at each site, the orbital wavefunction is given by a Gaussian envelope around a simple function that underpins the orbital symmetry, such that
\begin{align}\label{Eq:Supp_orbital}\tag{S.1}
    \psi_{i,\tau}(\vec{r}) = \mathcal{N}_{\tau} (x-\vec{R}_{i,x}-\bm{\tau}_{x})^{n}(y-\vec{R}_{i,y}-\bm{\tau}_{y})^{m}\text{e}^{-\gamma(\vec{r}-\vec{R}_i-\vec{d}_\tau)^2},
\end{align}
where $\mathcal{N}_\tau$ is a normalization factor, $\gamma$ describes the width of the Gaussian envelope, $\bm{\tau}$ is the position of the orbital $\tau$ in the $i^{th}$ unit cell (notice the use of \textit{boldface} to describe the position vector of the orbital), and $n,\, m$ parameterize the orbital. For example, for $s$-orbital $n=m=0$, for $p_x$ orbital $n=1,\, m=0 $, for $p_y$ orbital $n=0,\, m=1$. 
Generally, more complicated form factors can be used to consider other orbital symmetries. 
For our demonstration of $s$ and $p$ orbitals, the current form suffices.
To derive a tight binding Hamiltonian, we introduce the hopping amplitudes as the overlap integrals between localized orbitals
\begin{align}\label{Eq:Supp_hop}\tag{S.2}
    t_{i,\tau;j,\tau'} = \int d\vec{r} \psi_{i,\tau}(\vec{r}) \psi_{j,\tau'}(\vec{r}).
\end{align}
The hopping integrals are obtained  as
\begin{align}\label{Eq:Supp_hop_ex}\tag{S.3}
    t_{i,\tau;j,\tau'} \equiv t(\vec{u}_{i,\tau;j,\tau'}) = \begin{cases}
        t_{0} \text{e}^{-\frac{\gamma}{2}(u_x^2+u_y^2)} \quad \text{$\tau,\tau' = s$}\\
         - t_{0} u_y \text{e}^{-\frac{\gamma}{2}(u_x^2+u_y^2)} \quad \text{$\tau = s,\tau' = p_y$} \\
         - t_{0} u_{x}u_{y} \text{e}^{-\frac{\gamma}{2}(u_x^2+u_y^2)} \quad \text{$\tau = p_x,\tau' = p_y$} \\
         (t_{0} - t_1u^2_x) \text{e}^{-\frac{\gamma}{2}(u_x^2+u_y^2)} \quad \text{$\tau = p_x,\tau' = p_x$} \\
         (t_{0} - t_1 u^2_y) \text{e}^{-\frac{\gamma}{2}(u_x^2+u_y^2)} \quad \text{$\tau = p_y,\tau' = p_y$}
    \end{cases}.
\end{align}
Here $t_0 \equiv t_{0;\tau,\tau'}$ and $t_1 \equiv t_{1;\tau,\tau'}$ parameterize some Gaussian integrals and are independent of the displacement, $\vec{u} \equiv \vec{u}_{i,\tau;j,\tau'} = \vec{R}_j + \bm{\tau'} - \vec{R}_i - \bm{\tau}$. 
For brevity, we have omitted the subscripts in the displacement vector.
As the hopping depends on the displacement between the orbital position, the expansion around the symmetric lattice position of hopping parameter can be performed. The zeroth order term describes the standard electronic Hamiltonian
\begin{align}\label{Eq:Supp_Hee_zeroth}\tag{S.4}
    H_{e} = \sum_{\vec{R}_i}\sum_{\vec{R}_j} \sum_{\tau,\tau'} t(\vec{u}_0) c^{\dagger}_{\vec{R}_i+\vec{d}_{\tau}} c_{\vec{R}_j+\vec{d}_{\tau'}},
\end{align}
where $\vec{u}_0\equiv \vec{u}_{i\tau;j,\tau',0}$ describes the relative displacement between the hopping sites in absence of phonon displacements. 
We first describe the resultant electron hopping Hamiltonian for the two cases of the main text, \textit{i.e.}, (i) $a=1$, where $s$ and $p_y$ orbitals are involved and (ii) $a=2$, where $p_x$ and $p_y$ orbitals are involved.

For the case (i) [See Fig.~\ref{Fig:Orbitals_schematic} (a), (b), and (d)] the nearest neighbor hopping is inter-orbital, which is finite along $y$ direction and vanishes along $x$ direction. The next nearest neighbor hopping is intra-orbital. 
The resultant momentum space Hamiltonian takes the form
\begin{align}\label{Eq:Supp_Hee_case1}\tag{S.5}
    H^1_e = -\sum_{\vec{k}} \biggl [ 2t_s  \cos \frac{k_x}{\sqrt{2}} \cos \frac{k_y}{\sqrt{2}}  c^{\dagger}_s(\vec{k}) c_s(\vec{k}) + 2t_p  \cos \frac{k_x}{\sqrt{2}} \cos \frac{k_y}{\sqrt{2}}  c^{\dagger}_{p_y}c_{p_y} + 2it' \sin \frac{k_y}{\sqrt{2}} c^{\dagger}_sc_{p_y} + h.c \biggr ] .
\end{align}
Here on, we have taken the lattice constant to be unity.
Here, $t_s$ and $t_p$ parameterize the hopping, which have a more microscopic form given in Eq.~\ref{Eq:Supp_hop_ex}. 
The electron hopping Hamiltonian Eq.~\ref{Eq:Supp_Hee_case1} follows the form discussed in the main text. 

For the case (ii) [See Fig.~\ref{Fig:Orbitals_schematic} (a), (c), and (e)] the nearest neighbors hopping vanishes, the second nearest neighbor and the third nearest neighbor hoppings are intra-orbital. 
The fourth nearest neighbor hopping is inter-orbital and non-zero. 
For generality, we include upto the fourth nearest neighbor hopping to obtain 
\begin{align}\label{Eq:Supp_Hee_case2}
    & H^2_e = -\sum_{\vec{k}} \biggl [ \biggl ( 2t_{p1} \cos\frac{k_x}{\sqrt{2}}\cos\frac{k_y}{\sqrt{2}} + t_{p2} \cos \sqrt{2}k_x + t_{p3} \cos \sqrt{2}k_y \biggr ) c^{\dagger}_{p_y}(\vec{k}) c_{p_y}(\vec{k}) \notag\\
    &\hspace{4cm} + \biggl ( 2 t_{p1} \cos\frac{k_x}{\sqrt{2}}\cos\frac{k_y}{\sqrt{2}} +  t_{p2} \cos \sqrt{2}k_y +  t_{p3} \cos \sqrt{2} k_x \biggr ) c^{\dagger}_{p_x}(\vec{k}) c_{p_x}(\vec{k}) \notag\\
    &\hspace{4cm} +  4t' \biggl ( \sin \frac{k_x}{\sqrt{2}}\sin\sqrt{2}k_y + \sin \frac{k_y}{\sqrt{2}}\sin \sqrt{2}k_x \biggr ) c^{\dagger}_{p_x}(\vec{k}) c_{p_y}(\vec{k}) + h.c. \biggr ] \tag{S.6}
\end{align}
We can conclude that the symmetry based arguments of the main text are followed.

The first order term describes the typical electron-phonon coupling Hamiltonian
\begin{align}\label{Eq:supp_elec-pho}\tag{S.7}
    H_{e-ph} = \sum_{\vec{R}_i}\sum_{\vec{R}_j} \sum_{\tau,\tau'} d\vec{u} \cdot \nabla_{\vec{u}} t(\vec{u})|_{\vec{u}=\vec{u}_0} c^{\dagger}_{\vec{R}_i+\vec{d}_{\tau}} c_{\vec{R}_j+\vec{d}_{\tau'}},
\end{align}
where $d\vec{u} \equiv d\vec{u}_{i\tau;j,\tau'} = d\vec{R}_j+ d\bm{\tau}' - d\vec{R}_i - d\bm{\tau} = \vec{P}(\bar{\vec{r}}_{i,\tau;j,\tau'})/(Z_{eff}e)$ is parameterized by some local polarization per unit effective charge. 
Here we have assigned the local polarization vector to the average real space position, \textit{i.e.} $\bar{\vec{r}}_{i,\tau;j,\tau'} = (\vec{R}_i+\vec{R}_j+\bm{\tau}+\bm{\tau'})/2$.  
Notice for the hopping between identical orbitals or covalent bonds, the effective charge $Z_{eff}e\rightarrow 0$ and the polarization $\vec{P}\rightarrow 0$, the phonon displacement is well defined. 
At this point it is useful to write down the hoppings in the electron-phonon coupling in Eq.~\ref{Eq:supp_elec-pho} for our examples of $s$ and $p$ orbitals. 
\begin{align}\label{Eq:Supp_ele_ph_hop}\tag{S.8}
    d\vec{u}\cdot \nabla_{\vec{u}}t(\vec{u})|_{\vec{u}=\vec{u}_0}  = \begin{cases}
        -\gamma t_0 \text{e}^{-\frac{\gamma}{2}|u_0|^2}  (u_{0,x},\, u_{0,y} )\cdot \vec{P} \quad \text{$\tau,\tau' = s$}\\
         t_0 \text{e}^{-\frac{\gamma}{2}|u_0|^2} \biggl ( \gamma  u_{0,x}u_{0,y},\, \gamma u^2_{0,y} -1 \biggr )\cdot \vec{P} \quad \text{$\tau = s,\tau' = p_y$} \\
         -t_0 \text{e}^{-\frac{\gamma}{2}|u_0|^2} \biggl ( u_{0,y} -\gamma u^2_{0,x}u_{0,y},\,  u_{0,x}-\gamma u^2_{0,y}u_{0,x} \biggr )\cdot \vec{P} \quad \text{$\tau = p_x,\tau' = p_y$} \\
         t_0 \text{e}^{-\frac{\gamma}{2}|u_{0}|^2} \biggl ( (-\gamma t_0 -2t_1+\gamma t_1 u^2_{0,x})u_{0,x}, \, (-\gamma t_0 + t_1 \gamma u^2_{0,x})u_{0,y} \biggr )\cdot \vec{P} \quad \text{$\tau = p_x,\tau' = p_x$} \\
         t_0 \text{e}^{-\frac{\gamma}{2}|u_{0}|^2} \biggl ( (-\gamma t_0 + t_1 \gamma u^2_{0,y})u_{0,x},\, (-\gamma t_0 -2t_1+\gamma t_1 u^2_{0,y})u_{0,y}   \biggr )\cdot \vec{P} \quad \text{$\tau = p_y,\tau' = p_y$}
    \end{cases}.
\end{align}

After some algebra, for the case (i), we can obtain an electron-phonon Hamiltonian
\begin{align}\label{Eq:Supp_elec_ph_Ham1}\tag{S.9}
    H^1_{e-p} = -2t_0 \text{e}^{-\frac{\gamma}{8}} P_y(k'_x-k_x)\cos \frac{k_x+k'_x}{4} c^{\dagger}_{p_y}(\vec{k}') c_{s}(\vec{k}) +h.c.,
\end{align}
and for the case (ii), we obtain the electron-phonon Hamiltonian
\begin{align}\label{Eq:Supp_elec_ph_Ham2}\tag{S.10}
    H^2_{e-p} = it_0 \text{e}^{-\frac{\gamma}{8}}(-\gamma t_0 + \gamma t_1/4 ) P_x (k'_y-k_y) \sin \frac{k'_y+k_y}{4} c^{\dagger}_{p_y}(\vec{k}')c_{p_x}(\vec{k}) + h.c.
\end{align}
In obtaining Eqs.~\ref{Eq:Supp_elec_ph_Ham1} and~\ref{Eq:Supp_elec_ph_Ham2}, we have decomposed the optical phonon into the transverse and longitudinal brance and considered only the transverse branch.
For the small momenta we can expand the electron-phonon Hamiltonian over the power series of $\vec{k}$ and $\vec{k}'$. 
If we then only consider the first non-zero terms in the electron-phonon Hamiltonian, we have a form that follows Eqs.~\ref{Eq:Hq_couple_spin} and~\ref{Eq:Hq_couple} of the main text.

Having discussed some simple model, now we show that the TO phonon coupling considered for $\text{SrTiO}_2$ in Ref.~\cite{Gastiasoro2022,Yu2022} can be recast in the simple form discussed in this work. 
In $\text{SrTiO}_2$, $t_{2g}$ manifold of the $d$-orbitals are involved in the electronic Hamiltonian near the FS. These orbitals are even under parity operation and thus represent the case (ii) considered by us in the main text. Considering only two out of three $t_{2g}$ orbitals as done by Gastiasoro \textit{et. al.} in Ref~\cite{Gastiasoro2022},  the electron-phonon coupling is given by
\begin{align}\label{Eq:Supp_Hep_G}\tag{S.11}
    H_{e-p} = \frac{1}{2}\sum_{\vec{k},\vec{q}}\sum_{\lambda}\sum_{\mu,\nu} it' g_{\lambda,\vec{q}} c^{\dagger}_{\mu}(\vec{k}+\vec{q}) [\hat{\lambda}(\vec{q}) \times (2\vec{k}+\vec{q}) \cdot (\hat{\mu}\times\nu)] c_{\nu}(\vec{k})               
\end{align}
where $\lambda$ depicts the phonon mode and $\hat{\lambda}$ is a unit vector in the direction of phonon polarization and
\begin{align}\label{Eq:Supp_phonon_mode}\tag{S.12}
    g_{\lambda,\vec{q}} = \sqrt{\frac{\hbar}{2\mu_s\omega_{\lambda}(\vec{q})}} (a_{\lambda}(\vec{q})+a^{\dagger}_{\lambda}(-\vec{q}))
\end{align}
We can recast it as
\begin{align}\label{Eq:Supp_Hep_recast}\tag{S.13}
    H_{e-p} = -\frac{1}{2}\sum_{\vec{k},\vec{q}}\sum_{\lambda} t' g_{\lambda,\vec{q}} \hat{c}^{\dagger}(\vec{k}+\vec{q}) [\hat{\lambda}_i(\vec{q}) \{ (k_j+q_j) + k_j \} \epsilon_{ijz}]\tau_2 \hat{c}(\vec{k})
\end{align}
Finally by defining 
\begin{align}\label{Eq:Supp_couple_re}\tag{S.14}
    \Gamma^1_2(\vec{k}) = -k_y , \quad \Gamma^2_2 (\vec{k}) = k_x, \quad t'g_{\lambda,\vec{q}}\hat{\lambda}_i(\vec{q}) = \varphi_{\lambda,i}(\vec{q})
\end{align}
we can obtain Eq.~\ref{Eq:Hep_recast} in the form of Eq.~\ref{Eq:Hq_couple}, where $\varphi_{\lambda,i}$ is odd under inversion and the two orbitals have the same parity and thus $\Gamma(\vec{k}) = -\Gamma(-\vec{k})$. 

\subsection{Symmetry constraints on the S.O. coupling} 
Here we explicitly show that the $\mathcal{I}$ and $\mathcal{T}$ reversal symmetries constraint the SO coupling to the form of  Eqs.~\ref{Eq:SO_case1} and~\ref{Eq:SO_case2} of the main text.

Let's start with including a general SO coupling that is off-diagonal in spin  and written in the basis $\hat{c}(\vec{k}) = [c_{+,\uparrow}(\vec{k}),\, c_{-,\uparrow}(\vec{k}),\, c_{+,\downarrow}(\vec{k}),\, c_{-,\downarrow}(\vec{k}) ]^T$
\begin{align}\label{Eq:supp_T_cond1}\tag{S.15}
    H_{SO}(\vec{k}) = \begin{pmatrix}
     0 & 0 & \lambda_{0}(\vec{k})+\lambda_3(\vec{k}) & \lambda_1(\vec{k}) - i\lambda_2(\vec{k}) \\
     0 & 0 & \lambda_{1}(\vec{k})+i\lambda_2(\vec{k}) & \lambda_0(\vec{k}) - \lambda_3(\vec{k}) \\
     \lambda^{\ast}_0(\vec{k}) + \lambda^{\ast}_3(\vec{k}) 
     & \lambda^{\ast}_1(\vec{k}) -i \lambda^{\ast}_{2}(\vec{k}) & 0 & 0\\
     \lambda^{\ast}_1 (\vec{k}) + i\lambda^{\ast}_2(\vec{k}) & -\lambda^{\ast}_0(\vec{k})+\lambda^{\ast}_3(\vec{k}) & 0 & 0
     \end{pmatrix}
\end{align} 
Under the time reversal operation, the above Hamiltonian matrix transforms to
\begin{align}\label{Eq:supp_T_cond2}\tag{S.16}
    H_{SO}(\vec{k}) = \begin{pmatrix}
     0 & 0 & -\lambda_{0}(-\vec{k})-\lambda_3(-\vec{k}) & -\lambda_1(-\vec{k}) - i\lambda_2(-\vec{k}) \\
     0 & 0 & -\lambda_{1}(-\vec{k})+i\lambda_2(-\vec{k}) & -\lambda_0(-\vec{k}) + \lambda_3(-\vec{k}) \\
     -\lambda^{\ast}_0(-\vec{k}) - \lambda^{\ast}_3(-\vec{k}) 
     & -\lambda^{\ast}_1(-\vec{k}) -i \lambda^{\ast}_{2}(-\vec{k}) & 0 & 0\\
     -\lambda^{\ast}_1(-\vec{k}) + i\lambda^{\ast}_2(-\vec{k}) & -\lambda^{\ast}_0(-\vec{k})+\lambda^{\ast}_3(-\vec{k}) & 0 & 0
     \end{pmatrix}
\end{align}
If we consider the case (i) where the two orbitals have opposite sign under $\mathcal{I}$, we conclude that from $\mathcal{I}$, $\lambda_{0} (\vec{k})$ and $\lambda_{3} (\vec{k})$ are even functions of $\vec{k}$, while $\lambda_{1}(\vec{k})$ and   $\lambda_{2}(\vec{k})$ are odd functions of $\vec{k}$. 
Under the $\mathcal{T}$ symmetry the matrix in Eq.~\ref{Eq:supp_T_cond1} and Eq.~\ref{Eq:supp_T_cond2} must be equal. Thus we obtain that only non-zero term that is allowed is $\lambda_1(\vec{k})$. 

Similarly, for case (ii) where the two orbitals have same sign under $\mathcal{I}$, all $\lambda_{i}(\vec{k})$ must be even functions of $\vec{k}$ from the $\mathcal{I}$ symmetry. 
Thus $\lambda_2(\vec{k})$ remains the only non-zero term when the $\mathcal{T}$ is simultaneously satisfied.

\subsection{Gap equations}
For a spin-orbit coupled system, the general multi-component linearized gap equation near $T_c$ can be expressed as~\cite{Sigrist1991}
\begin{align}\label{Eq:Supp_}\tag{S.17}
    \Delta^{\beta,\alpha}_{s,s'} (\vec{k}) = -k_B T_c \sum_{n} \sum_{\vec{k}'} \sum_{\gamma,\delta,\eta,\nu} \sum_{s_1,s_2,s_3,s_4} V^{\alpha,\beta,\gamma,\delta}_{s',s,s_1,s_2} (\vec{k},\vec{k}') G^0_{\gamma,\eta;s_1,s_3} (\vec{k}',i\omega_n) G^0_{\delta,\nu;s_2,s_4} (-\vec{k}',-i\omega_n) \Delta^{\eta,\nu}_{s_3,s_4}(\vec{k}'),
\end{align}
where $\alpha, \beta,\gamma,\delta,\eta,\nu$ are the atomic orbital labels and $s,s',s_1,s_2,s_3,s_4$ are the spin labels. 
For our case in the appropriated spin rotation basis, the single particle Green function $G^0$ are diagonal in spin. 
Moreover, because of the spin conserving electron-phonon scatterings the interaction matrix elements are $ V^{\alpha,\beta,\gamma,\delta}_{s',s,s_1,s_2} = V_{\alpha,\beta,\gamma,\delta}\delta_{s',s_2}\delta_{s,s_1}$. 
We obtain the simplification
\begin{align}\tag{S.18}
    \Delta^{\beta,\alpha}_{s,s'} (\vec{k}) = -k_B T_c \sum_{n} \sum_{\vec{k}'} \sum_{\gamma,\delta,\eta,\nu}  V_{\alpha,\beta,\gamma,\delta} (\vec{k},\vec{k}') G^0_{\gamma,\eta;s} (\vec{k}',i\omega_n) G^0_{\delta,\nu;s'} (-\vec{k}',-i\omega_n) \Delta^{\eta,\nu}_{s,s'}(\vec{k}').
\end{align}
Thus the same spin and opposite spin correlations do not mix. 
However, within the opposite spin correlation sector, we can consider spin singlet(triplet) gaps by taking the liner combinations $\hat{\Delta}_{\uparrow,\downarrow} \mp \hat{\Delta}_{\downarrow,\uparrow}$ respectively. 
Here $\hat{.}$ represents orbital space. 
We can write down linearized gap equations for spin-singlet and spin-triplet pairing channels.
\begin{align}\label{Eq:Supp_singlet_triplet}
    & \Delta^{\beta,\alpha}_{\text{singlet}/\text{triplet}} (\vec{k}) = -k_B T_c \sum_{n} \sum_{\vec{k}'} \sum_{\gamma,\delta,\eta,\nu}  V_{\alpha,\beta,\gamma,\delta} (\vec{k},\vec{k}') [G^0_{\gamma,\eta;\uparrow} (\vec{k}',i\omega_n) G^0_{\delta,\nu;\downarrow} (-\vec{k}',-i\omega_n) \Delta^{\eta,\nu}_{\uparrow,\downarrow}(\vec{k}') \notag\\
    & \hspace{8cm}\mp G^0_{\gamma,\eta;\downarrow} (\vec{k}',i\omega_n) G^0_{\delta,\nu;\uparrow} (-\vec{k}',-i\omega_n) \Delta^{\eta,\nu}_{\downarrow,\uparrow}(\vec{k}')].\tag{S.19}
\end{align}
In the presence of the SO coupling, since $G^{0}_{\uparrow} \neq  G^{0}_{\downarrow}$, even if the two spins form two degenerate FS. In particular, upto first order in the SO coupling, the single particle Green function can be decomposed over the spin independent Green function and SO Green function $G^0_s = g^0+sg^{SO}$ and the singlet/triplet gap equations reduce to 
\begin{align}\label{Eq:Supp_singlet_triplet1}
    & \Delta^{\beta,\alpha}_{\text{singlet}/\text{triplet}} (\vec{k}) \sim -k_B T_c \sum_{n} \sum_{\vec{k}'} \sum_{\gamma,\delta,\eta,\nu}  V_{\alpha,\beta,\gamma,\delta} (\vec{k},\vec{k}') [g^0_{\gamma,\eta} (\vec{k}',i\omega_n) g^0_{\delta,\nu} (-\vec{k}',-i\omega_n) \Delta^{\eta,\nu}_{\text{singlet/triplet}}(\vec{k}') \notag\\
    &\hspace{3cm} + \{ g^{SO}_{\gamma,\eta}(\vec{k}',i\omega_n)g^{O}_{\delta,\nu}(-\vec{k}',-i\omega_n)+ g^0_{\gamma,\eta}(\vec{k}',i\omega_n) g^{SO}_{\delta,\nu}(-(\vec{k}',-i\omega_n)\} \Delta_{\text{triplet/singlet}}].\tag{S.20}
\end{align}
We see that the singlet-triplet gap equations are not decoupled in the presence of the SO coupling. 

\subsection{Pairing in the band basis}
Here, starting from the orbital-spin basis, we transform the electron-phonon Hamiltonian and the resultant pairing order parameter to the band basis. 
Since in the appropriate spin basis, the two spin sectors are decoupled, in any particular spin $`s'$, the single electron Hamiltonian is expanded over the Pauli matrices $\tau$ in the orbital basis (See Eq.~\ref{Eq:H_elec} in the main text).
 We diagonalize the electron part of the BdG (upper diagonal block) as $H_{e,s} = \sum_{b}\sum_{\vec{k}} \chi^{\dagger}_{s,b}(\vec{k}) E_{s,b} (\vec{k}) \chi_{s,b}(\vec{k})$, where $b = l,u$ label the lower and upper electron bands associated with spin $s$, such that
\begin{align}\label{Eq:Supp_band}\tag{S.21}
    E_{s,l/b}(\vec{k}) = h_{0,s}(\vec{k}) \mp |h_{s}(\vec{k})|,
\end{align}
where
\begin{align}\label{Eq:Supp_band1}\tag{S.22}
   |h_{s}(\vec{k})|=  \sqrt{\sum^3_{i=1}|h_{i,s}(\vec{k})|^2}
\end{align}
and the standard diagonalization is done under the transformation $\hat{\chi}_{s}(\vec{k}) = T_{s} \hat{c}_s(\vec{k})$, where
\begin{align}\label{Eq:Supp_trans}\tag{S.23}
    T_{s}(\vec{k}) = \begin{pmatrix}
    -\sin\frac{\theta_{\vec{k}}}{2} & \text{e}^{-i\phi_{s,\vec{k}}}\cos \frac{\theta_\vec{k}}{2} \\
    \text{e}^{i\phi_{s,\vec{k}}}\cos \frac{\theta_\vec{k}}{2} & \sin\frac{\theta_{\vec{k}}}{2}
    \end{pmatrix}_s,
\end{align}
and
\begin{align}\label{Eq:supp_diag_param1}\tag{S.24}
    \sin\frac{\theta_{\vec{k},s}}{2} = \sqrt{\frac{|h_{s}(\vec{k})|-h_{3,s}(\vec{k})}{2|h_{s}(\vec{k})|}}, \quad 
    \cos\frac{\theta_{\vec{k},s}}{2} = \sqrt{\frac{|h_{s}(\vec{k})|+h_{3,s}(\vec{k})}{2|h_{s}(\vec{k})|}}, \quad \text{e}^{-i\phi_{s,\vec{k}}} = \frac{h_{1,s}(\vec{k}) - i h_{2,s}(\vec{k})}{|h_s(\vec{k})|-|h_3(\vec{k})}| .
\end{align}
We notice that the first two parameters in Eq.~\ref{Eq:supp_diag_param1} are even functions of $\vec{k}$ and do not depend on the spin $s$, irrespective of nature of the atomic orbitals. 
The third parameter depends on the spin and the nature of the involved atomic orbitals. 
More precisely, the third parameter in Eq.~\ref{Eq:supp_diag_param1} is odd (even) function in $\vec{k}$ for the case $a=1(2)$ and $\text{e}^{-i\phi_{s,\vec{k}}}=\text{e}^{i\phi_{-s,\vec{k}}}$.

Under this procedure the electron-phonon Hamiltonian is transformed as follows:
\begin{align}\label{Eq:Supp_ep_transformation}\tag{S.25}
    H^{a}_{e-p} = \frac{1}{2}\sum_{\vec{k},\vec{k}'}\sum_s \varphi(\vec{k}'-\vec{k})\hat{\chi}^{\dagger}_s(\vec{k}') \biggl ( [\Gamma_{a;1,s}(\vec{k},\vec{k'})]U^{\dagger}_s(\vec{k}')\tau_1 U_{s}(\vec{k})+ [\Gamma_{a;2,s}(\vec{k},\vec{k}')] U^{\dagger}_s(\vec{k}')\tau_1 U_{s}(\vec{k}) \biggr ) \hat{\chi}_s(\vec{k}).
\end{align}
Here, we have used subscripts $a$ and $s$ in the vertex functions, so that the two different cases described by Eqs.~\ref{Eq:Hq_couple_spin} and~\ref{Eq:Hq_couple} of the main text can be expressed in one representation. 
Explicitly carrying out the transformation in the large brackets in Eq.~\ref{Eq:Supp_ep_transformation}, we obtain Eq.~\ref{Eq:Ham_e_ph_band} of the main text, where
\begin{subequations}\label{Eq:Supp_vertex_funtion_band}
\begin{align}
     & \bar{\Gamma}^a_{0,s}(\vec{k},\vec{k}') = i\biggl ( \sin\phi_{\vec{k}'}\cos\frac{\theta_{\vec{k}'}}{2} \sin\frac{\theta_{\vec{k}}}{2} - \sin\phi_{\vec{k}}\cos\frac{\theta_{\vec{k}}}{2} \sin\frac{\theta_{\vec{k}'}}{2} \biggr )\Gamma_{a;1}(\vec{k},\vec{k}') \notag\\
     &\hspace{6cm} +  
     i\biggl (\cos\phi_{\vec{k}}\cos\frac{\theta_{\vec{k}}}{2} \sin\frac{\theta_{\vec{k}'}}{2} -\cos\phi_{\vec{k}'}\sin\frac{\theta_{\vec{k}}}{2} \cos\frac{\theta_{\vec{k}'}}{2} \biggr )\Gamma_{a;2}(\vec{k},\vec{k}'), \tag{S.26a}\\
     & \bar{\Gamma}^a_{1,s}(\vec{k},\vec{k}') = s^{a-1}\biggl ( 
     \cos\phi_{\vec{k}}\cos\frac{\theta_{\vec{k}}}{2}\cos\phi_{\vec{k}'}\cos\frac{\theta_{\vec{k}'}}{2}
     -\sin\phi_{\vec{k}}\cos\frac{\theta_{\vec{k}}}{2}\sin\phi_{\vec{k}'}\cos\frac{\theta_{\vec{k}'}}{2}
     -\sin\frac{\theta_{\vec{k}}}{2}\sin\frac{\theta_{\vec{k}'}}{2}
     \biggr )\Gamma_{a;1}(\vec{k},\vec{k}') \notag\\
     &\hspace{4cm} + s^{a-1} 
     \biggl (\sin\phi_{\vec{k}}\cos\frac{\theta_{\vec{k}}}{2}\cos\phi_{\vec{k}'}\cos\frac{\theta_{\vec{k}'}}{2}
     +\cos\phi_{\vec{k}}\cos\frac{\theta_{\vec{k}}}{2}\sin\phi_{\vec{k}'}\cos\frac{\theta_{\vec{k}'}}{2}     
      \biggr )\Gamma_{a;2}(\vec{k},\vec{k}'), \tag{S.26b}\\
      & \bar{\Gamma}^a_{2,s}(\vec{k},\vec{k}') = s^a\biggl ( \sin\phi_{\vec{k}}\cos\frac{\theta_{\vec{k}}}{2}\sin\phi_{\vec{k}'}\cos\frac{\theta_{\vec{k}'}}{2}
     -\cos\phi_{\vec{k}}\cos\frac{\theta_{\vec{k}}}{2}\cos\phi_{\vec{k}'}\cos\frac{\theta_{\vec{k}'}}{2}
     -\sin\frac{\theta_{\vec{k}}}{2}\sin\frac{\theta_{\vec{k}'}}{2}
     \biggr )\Gamma_{a;2}(\vec{k},\vec{k}') \notag\\
     &\hspace{4cm} +  
     s^a \biggl (\sin\phi_{\vec{k}}\cos\frac{\theta_{\vec{k}}}{2}\cos\phi_{\vec{k}'}\cos\frac{\theta_{\vec{k}'}}{2}
     +\cos\phi_{\vec{k}}\cos\frac{\theta_{\vec{k}}}{2}\sin\phi_{\vec{k}'}\cos\frac{\theta_{\vec{k}'}}{2}     
      \biggr )\Gamma_{a;1}(\vec{k},\vec{k}'), \tag{S.26c}\\
    & \bar{\Gamma}^a_{3,s}(\vec{k},\vec{k}') = -s\biggl ( \cos\phi_{\vec{k}'}\cos\frac{\theta_{\vec{k}}}{2} \sin\frac{\theta_{\vec{k}'}}{2} + \cos\phi_{\vec{k}'}\sin\frac{\theta_{\vec{k}}}{2} \cos\frac{\theta_{\vec{k}'}}{2} \biggr )\Gamma_{a;1}(\vec{k},\vec{k}') \notag\\
     &\hspace{6cm} -s  
     \biggl (\sin\phi_{\vec{k}}\cos\frac{\theta_{\vec{k}}}{2} \sin\frac{\theta_{\vec{k}'}}{2} + \sin\phi_{\vec{k}'}\sin\frac{\theta_{\vec{k}}}{2} \cos\frac{\theta_{\vec{k}'}}{2} \biggr )\Gamma_{a;2}(\vec{k},\vec{k}').\tag{S.26d} 
\end{align}
\end{subequations}
We notice that the $\mathcal{T}$ symmetry constrain in Eq.~\ref{Eq:elec-pho-vertex-T} of the main text is followed. 
We can also verify the $\mathcal{I}$-odd condition is followed as 
\begin{subequations}\label{Eq:Supp_inversion}
\begin{align}
    &\bar{\Gamma}^1_{j,s}(-\vec{k},-\vec{k}')  = \begin{cases}
     -\bar{\Gamma}^1_{j,s}(-\vec{k},-\vec{k}') \quad \text{$j = 0,3$}  \tag{S.27a}\\
     \bar{\Gamma}^1_{j,s}(-\vec{k},-\vec{k}') \quad \text{$j = 1,2$} 
    \end{cases}\\
    & \bar{\Gamma}^2_{j,s}(-\vec{k},-\vec{k}')  = - \bar{\Gamma}^2_{j,s}(\vec{k},\vec{k}') .\tag{S.27b}
\end{align}
\end{subequations}

In same spin channel, since the spin part of the pairing order parameter is triplet, for the overall anti-symmetry, we the $2\times 2$ pairing order parameter in the orbital basis
\begin{align}\label{Eq:Supp_2by2gap}
    \Delta_{s,s}(\vec{k}) = i[\vec{d}(\vec{k})\cdot \bm{\tau}+D(\vec{k}) ]\tau_2,\tag{S.28}
\end{align}
where $D(\vec{k}) = D(-\vec{k})$ and $\vec{d}(\vec{k}) = -\vec{d}(-\vec{k})$ are even and odd functions of $\vec{k}$ respectively. 
Transforming the pairing order parameter to the band basis and only considering the diagonal elements, we obtain
\begin{align}\label{Eq:Supp_gap_band}
    \Delta_{s,s}(\vec{k}) = \begin{pmatrix}
        \Delta_{l;s,s}(\vec{k}) & 0\\
        0 & \Delta_{u;s,s}(\vec{k})
    \end{pmatrix},\tag{S.29}
\end{align}
where
\begin{subequations}\label{Eq:Supp_gap_band1}
\begin{align}
    & \Delta_{l;s,s}(\vec{k}) = \sin^2\frac{\theta_{\vec{k}}}{2} [d_0(\vec{k})+d_3(\vec{k})] + 
    \text{e}^{-2i\phi_{\vec{k}}} \cos^2\frac{\theta_{\vec{k}}}{2} [d_0(\vec{k})- d_3(\vec{k})] - \text{e}^{-i\phi_{\vec{k}}}  \sin\theta_{\vec{k}} d_1(\vec{k}), \tag{S.30a}\\
    & \Delta_{u;s,s}(\vec{k}) = \sin^2\frac{\theta_{\vec{k}}}{2} [d_0(\vec{k})-d_3(\vec{k})] + 
    \text{e}^{2i\phi_{\vec{k}}} \cos^2\frac{\theta_{\vec{k}}}{2} [d_0(\vec{k}) + d_3(\vec{k})] + \text{e}^{i\phi_{\vec{k}}}  \sin\theta_{\vec{k}} d_1(\vec{k}).\tag{S.30b}
\end{align}
\end{subequations}

Since for the case (ii), $h_1$ and $h_2$ are even functions of $\vec{k}$ because of the $\mathcal{I}$-symmetry, the exponential factors are also even in $\vec{k}$. Thus each term on the R.H.S. above is a product of an off function and an even function in $\vec{k}$. 
Thus overall, the R.H.S. is an odd-function of $\vec{k}$, leading to an odd-parity gap. 

For the case (i), however, $h_1$ and $h_2$ are odd functions of $\vec{k}$ and as a result, the exponential factors are also odd in $\vec{k}$. 
Thus while the last term on the R.H.S. is am even function in $\vec{k}$ ( product of two odd functions in $\vec{k}$), while the first two terms on the R.H.S. still remain odd in $\vec{k}$. 
Thus overall, the band projected gap functions in the same spin channels are mixed parity. 

The matrix elements of the projection of the Hubbard interaction on the band basis are given by
\begin{align}
    \bar{U}^{a}_{s}(\vec{k},\vec{k}') = \sum_{\tau,\tau'} \sqrt{U} T^a_{s,\tau,\nu}(\vec{k}')T^{a\ast}_{s,\tau',\nu}(\vec{k}),\tag{S.31}
\end{align}
where $T^{a}_{s,\tau,\nu} (\vec{k})$ are the matrix elements of the diagonalization transformation.

\end{widetext}

\end{document}